\documentclass{aa}

\usepackage{amsmath}
\usepackage{amssymb}

\usepackage{graphicx, xcolor}
\usepackage{txfonts}
\usepackage[linkcolor=red,citecolor=blue,filecolor=cyan,urlcolor=magenta, colorlinks=true, allbordercolors={white}]{hyperref}
\def\Oiii{[O\,{\sc iii}]}   \def\Ci{[C\,{\sc i}]}
  \def\Cii{[C\,{\sc ii}]}
\def\Ci{[C\,{\sc i}]}    
 \def\aco{{\rm CO}(1-0)}

   \def\lsim{\mathrel{\rlap{\lower
3pt \hbox{$\sim$}} \raise 2.0pt \hbox{$<$}}} \def\gsim{\mathrel{\rlap{\lower 3pt \hbox{$\sim$}}
\raise 2.0pt \hbox{$>$}}}

\begin{document}

\title{The ALMA Reionization Era Bright Emission Line Survey}
\subtitle{The molecular gas content of galaxies at $z\sim7$}

\author{
M. Aravena\inst{1}
\and
K. Heintz\inst{2,3}
\and
M. Dessauges-Zavadsky\inst{4}
\and
P. Oesch\inst{2,3,4}
\and
H. Algera\inst{5,6}
\and
R. Bouwens\inst{7}
\and
E. Da Cunha\inst{8}
\and
P. Dayal\inst{9}
\and
I. De Looze\inst{10}
\and
A. Ferrara\inst{11}
\and
Y. Fudamoto\inst{6,12}
\and
V. Gonzalez\inst{13}
\and
L. Graziani\inst{14,15}
\and
H. Inami\inst{5}
\and
A. Pallottini\inst{11}
\and
R. Schneider\inst{14,15,16,17}
\and
S. Schouws\inst{7}
\and
L. Sommovigo\inst{14}
\and
M. Topping\inst{18}
\and
P. van der Werf\inst{7}
\and
M. Palla\inst{10}
}
\institute{
Instituto de Estudios Astrof\'{\i}sicos, Facultad de Ingenier\'{\i}a y Ciencias, Universidad Diego Portales, Av. Ej\'ercito 441, Santiago 8370191, Chile\\ 
\email{manuel.aravenaa@mail.udp.cl}
\and
Cosmic Dawn Center (DAWN), Denmark
\and
Niels Bohr Institute, University of Copenhagen, Jagtvej 128, DK- 2200, Copenhagen N, Denmark
\and
Department of Astronomy, University of Geneva, Chemin Pegasi 51, 1290 Versoix, Switzerland
\and
Hiroshima Astrophysical Science Center, Hiroshima University, 1-3-1 Kagamiyama, Higashi-Hiroshima, Hiroshima 739-8526, Japan
\and
National Astronomical Observatory of Japan, 2-21-1, Osawa, Mitaka, Tokyo, Japan
\and
Leiden Observatory, Leiden University, NL-2300 RA Leiden, The Netherlands
\and
International Centre for Radio Astronomy Research, University of Western Australia, 35 Stirling Hwy., Crawley, WA 6009, Australia
\and
Kapteyn Astronomical Institute, University of Groningen, P.O. Box 800, 9700 AV Groningen, The Netherlands
\and
Sterrenkundig Observatorium, Ghent University, Krijgslaan 281-S9, B-9000 Gent, Belgium
\and
Scuola Normale Superiore, Piazza dei Cavalieri 7, I-56126 Pisa, Italy
\and
Waseda Research Institute for Science and Engineering, Faculty of Science and Engineering, Waseda University, 3-4-1 Okubo, Shinjuku, Tokyo 169-8555, Japan
\and
Departmento de Astronomia,Universidad de Chile, Casilla 36-D, Santiago 7591245, Chile
\and
Dipartimento di Fisica, Sapienza, Universita di Roma,Piazza le Aldo Moro 5, I-00185 Roma, Italy
\and
INAF/Osservatorio Astronomico di Roma, via Frascati 33, I 00078 Monte Porzio Catone, Roma, Italy
\and
Sapienza School for Advanced Studies, Sapienza Università di Roma, P.le Aldo Moro 2, 00185 Roma, Italy
\and
INFN, Sezione di Roma 1, P.le Aldo Moro 2, 00185 Roma, Italy
\and
Steward Observatory, University of Arizona, 933 N Cherry Ave, Tucson, AZ 85721 USA
\\
}

\date{Received November 15, 2022; accepted November 16, 2022}

\authorrunning{Aravena et~al.}
\titlerunning{Molecular gas content in galaxies at $z\sim7$}

 
  \abstract{A key to understanding the formation of the first galaxies is to quantify the content of the molecular gas as the fuel for star formation activity through the epoch of reionization. In this paper, we use the 158$\mu$m \Cii{} fine-structure emission line as a tracer of the molecular gas in the interstellar medium (ISM) in a sample of $z=6.5-7.5$ galaxies recently unveiled by the  Reionization Era Bright Line Emission Survey, REBELS, with the Atacama Large Millimeter/submillimeter Array. We find substantial amounts of molecular gas ($\sim10^{10.5}\ M_\sun$) comparable to those found in lower redshift galaxies for similar stellar masses ($\sim10^{10}\ M_\sun$). The REBELS galaxies appear to follow the  standard scaling relations of molecular gas to stellar mass ratio ($\mu_{\rm mol}$) and gas depletion timescale ($t_{\rm dep}$) with distance to the star-forming main-sequence expected from extrapolations of $z\sim1-4$ observations. We find median values at $z\sim7$ of $\mu_{\rm mol}=2.6_{-1.4}^{4.1}$ and $t_{\rm dep}=0.5_{-0.14}^{+0.26}$ Gyr, indicating that the baryonic content of these galaxies is gas-phase dominated and little evolution from $z\sim7$ to 4. Our measurements of the cosmic density of molecular gas, log$(\rho_{\rm mol}/(M_\sun {\rm Mpc}^{-3}))=6.34^{+0.34}_{-0.31}$ , indicate a steady increase by an order of magnitude from $z\sim7$ to 4.} 


\maketitle
%

\section{Introduction} \label{sec:intro}

One of the most important advances in astrophysics in the past two decades has been the determination of the evolution of the cosmic star formation rate (SFR) density from cosmic dawn to the present time. This quantity was found to rapidly rise from $z=10$ to $z=3$ (a span of 3 Gyrs of cosmic time), reaching a plateau at $z=1-3$, and then steadily declining at $z<1$ \citep[e.g.,][]{madau14, peroux20}. The cosmic SFR activity appears dominated by star-forming galaxies that grow secularly, forming a tight correlation in the SFR versus stellar mass plane, usually termed as the star-forming ``main-sequence'' \citep[MS;][]{brinchmann08, daddi07, elbaz07, elbaz11, noeske07, peng10, rodighiero10, whitaker10, whitaker14, speagle14, schreiber15, iyer18, dicesare23}. These galaxies sporadically will be involved in major galaxy interactions or mergers, leading to increased SFRs \citep[e.g.;][]{kartaltepe12}, and eventually will halt star formation, possibly through exhaustion of their cold gas reservoirs that sustain star formation \citep[]{peng10, sargent14, spilker18, belli21, williams21, bezanson19, bezanson22}. 

This framework for galaxy growth has been built through intensive multiwavelength observational campaigns and simulations. Determination of the molecular gas content has been paramount, as the cold gas represents the fuel for star-forming activity. Since directly observing $H_2$ molecular gas is difficult, measurements of the molecular gas mass in distant galaxies have focused on the dust continuum and the CO and \Ci{} line emission \citep[e.g.,][]{heintz20}. Dedicated targeted surveys and archival searches for such dust and CO/\Ci{} line observations have led to thousands of galaxies with measured molecular gas masses \citep[e.g.;][]{daddi08, daddi10, tacconi13, tacconi18, dessauges15, dessauges17, freundlich19, scoville14, scoville17, fudamoto17, liu19, valentino18, valentino20}. A complementary approach has been performing deep systematic observations of dust continuum and CO line emission in contiguous patches of the sky centered on cosmological deep fields \citep{carilli02, decarli14, walter14, walter16, aravena16b, dunlop17, pavesi18, riechers19, franco20, hatsukade18, gonzalezlopez19, gonzalezlopez20, decarli20}. Such deep field surveys have the advantage of avoiding pre-selection biases compared to targeted surveys, although by design they are restricted to fainter objects. Overall, these programs have enabled establishing a set of scaling relations that allow us to describe the galaxy growth in terms of the galaxies' stellar mass, specific SFRs, and redshift, out to $z\sim4$. In particular, current studies indicate that from $z=3$ to 1, there is an increase in the molecular gas depletion timescales, a decrease in the molecular gas fractions, and a decrease of the specific SFR. Also, the molecular gas depletion timescales decrease with increasing specific SFR \citep[e.g.; ][]{tacconi20}. Of particular interest has been the determination of the cosmic density of molecular gas ($\rho_{\rm H2}$) as a function of redshift from various dust continuum and CO line surveys out to $z\sim5$ \citep{walter14, decarli16a, decarli19, decarli20, riechers19, magnelli20}. This quantity has been measured to closely follow the evolution of the cosmic density of SFR, supporting a relatively unchanged star formation efficiency on cosmic scales. Based on these results, it has also become evident that gas accretion from the intergalactic medium (IGM) is necessary to sustain the build-up of stellar mass in galaxies, at least since $z=1.5$ \citep{walter20}.

Due to cosmic dimming, the decreasing metallicities at higher redshifts, which make the dust continuum and CO/\Ci line emission fainter, and the increase of the cosmic microwave background with redshift, it becomes increasingly difficult to measure the molecular gas reservoirs of galaxies at $z>4$ \citep[e.g.;][]{carilli13}. 
Since the \Cii{} line is among the brightest coolant lines in galaxies, and because it is redshifted into the millimeter atmospheric windows observable from the ground at $z>4$, it has thus become a powerful tool to measure the interstellar medium (ISM) properties of galaxies. Even though its emission arises from various environments, including the neutral, ionized, and molecular gas phases \citep[e.g.;][]{vallini15, pallotini17,lagache18, olsen17}, it has been reliably calibrated and used as a molecular gas tracer for massive galaxies at $z<0.2$ \citep{hughes17, madden20} and at $z>2$ \citep{zanella18, vizgan22}. While \Cii{} has been traditionally used as a tracer of star formation in galaxies, the close link between star formation and molecular gas content (i.e. the Schmidt-Kennicutt relation) yields naturally a relation between \Cii{} and molecular gas. Current studies show that \Cii{} would even trace the molecular gas better than CO for low-metallicity environments where the gas is CO-dark, which is the case for high-redshift galaxies \citep{madden20, vizgan22}.

The brightness of the \Cii{} line has allowed for the molecular gas measurements of galaxies at z$\sim4-6$ from the Atacama Large Millimeter/submillimeter Array (ALMA) Large Program to INvestigate C$+$ at Early Times (ALPINE) survey \citep{dessauges20}.

Here, we aim to extend such exploration to even higher redshifts, by using data from the Reionization Era Bright Emission Line Survey \citep[REBELS;][]{bouwens22}. REBELS is an ALMA large program aimed at confirming the redshifts of a sample of 40 UV-bright star-forming galaxy candidates at $z\sim6-9$ using \Cii{} 158 $\mu$m and \Oiii{} 88$\mu$m emission line scans of the high-probability redshift range obtained from photometric redshift measurements. REBELS has enabled studies of the properties of these early galaxies through simultaneous studies of the dust continuum and \Cii{} emission line, including: determination of the dust properties \citep{inami22, sommovigo22, ferrara22, dayal22}, the IR luminosity function \citep{barrufet23} and obscured SFR density and fractions at $z\sim7$ \citep{algera23a}, measurements of the dust temperature and [OIII]]/[CII] line ratios in a few systems \citep{algera23b}, specific SFRs \citep{topping22}, the discovery of inconspicuous overdensities of dusty star-forming galaxies \citep{fudamoto21}, measurements of cosmic HI gas mass densities \citep{heintz22}, [CII] sizes \citep{fudamoto22} identification of Lyman-$\alpha$ emission and velocity offsets \citep{endsley22}, and a resolved study of a massive star-forming system at $z=7.3$ \citep{hygate23}.

In this paper, we present molecular gas mass estimates for a sample of 28 \Cii-detected star-forming galaxies at $z\sim6-8$ drawn from REBELS \citep{bouwens22} and its pilot surveys \citep{smit18, schouws22a, schouws22b}. These measurements are thereby used to provide estimates of the evolution of the molecular gas depletion timescales, molecular gas ratios, and cosmic molecular gas density at $z\sim7$. In Section \ref{sec:data}, we present details about the REBELS survey, the available multi-wavelength data, and comparison samples used throughout this work. In Section \ref{sec:gasestimates}, we discuss the use of the \Cii-based molecular gas calibration in the REBELS sample, comparing it to alternative estimates of the molecular gas content available. In Section \ref{sec:analysis}, we present measurements of the molecular gas depletion timescales, gas fraction, and cosmic molecular gas density, and discuss them in the context of the established scaling relations and predicted evolution of these quantities out to $z\sim7$. In Section \ref{sec:conclusions} we provide a summary of the main results of this work. Throughout the paper, we adopt a standard $\Lambda$CDM cosmology\footnote{$H_0$ = 70 km s$^{-1}$ Mpc$^{-1}$, $\Omega_{\rm M}= 0.3$ and $\Omega_\Lambda=0.7$}, and a \citet{chabrier03} initial mass function (IMF), with stellar masses ranging from $0.1-300\ M_\sun$.

\section{Data}
\label{sec:data}

\subsection{Observations and sample}

Our study takes advantage of the \Cii{} line emission observations in a sample of $z\sim6-8$ galaxies obtained by the ALMA REBELS large program (PID: 2019.1.01634.L). The REBELS survey details are provided in \citet{bouwens22} and Schouws et al. (in preparation). Here, we provide a brief description of the observations. 

The REBELS program used line scans to search for redshifted \Cii{} 158$\mu$m and \Oiii 88$\mu$m lines from sources that had previously been selected as candidate $z>6$ galaxies based on their UV luminosity ($M_{\rm UV}<-21.5$ mag) and photometric redshift probability distributions ($6.5<z<9.5$), following an approach similar to the pilot programs \citep{smit18, schouws22a}. 


The REBELS survey targeted 40 galaxies selected from a total area of 7 deg$^2$ encompassing various deep fields (Sec. \ref{sec:ancillary}). Line scans searching for the \Cii \ and \Oiii \ lines were conducted for 36 galaxies at $z\sim6.5-7.5$ and four galaxies at $z\sim8.0-9.5$ using ALMA bands 6 and 8, respectively. For each target, the scanning range was chosen to cover 90\% of the relevant frequency range following the photometric redshift probability distribution. 

The ALMA observations used in this study were performed in the most compact configurations, C43-1 and C43-2, yielding synthesized beam sizes of $1.2-1.6''$ in band 6. The [CII] line scans were performed to a sensitivity of $2\times10^8 L_\sun$ at $z=7$ \citep[$5\sigma$;][]{bouwens22}.

\begin{figure*}[ht]
    \centering
    \includegraphics[scale=0.6]{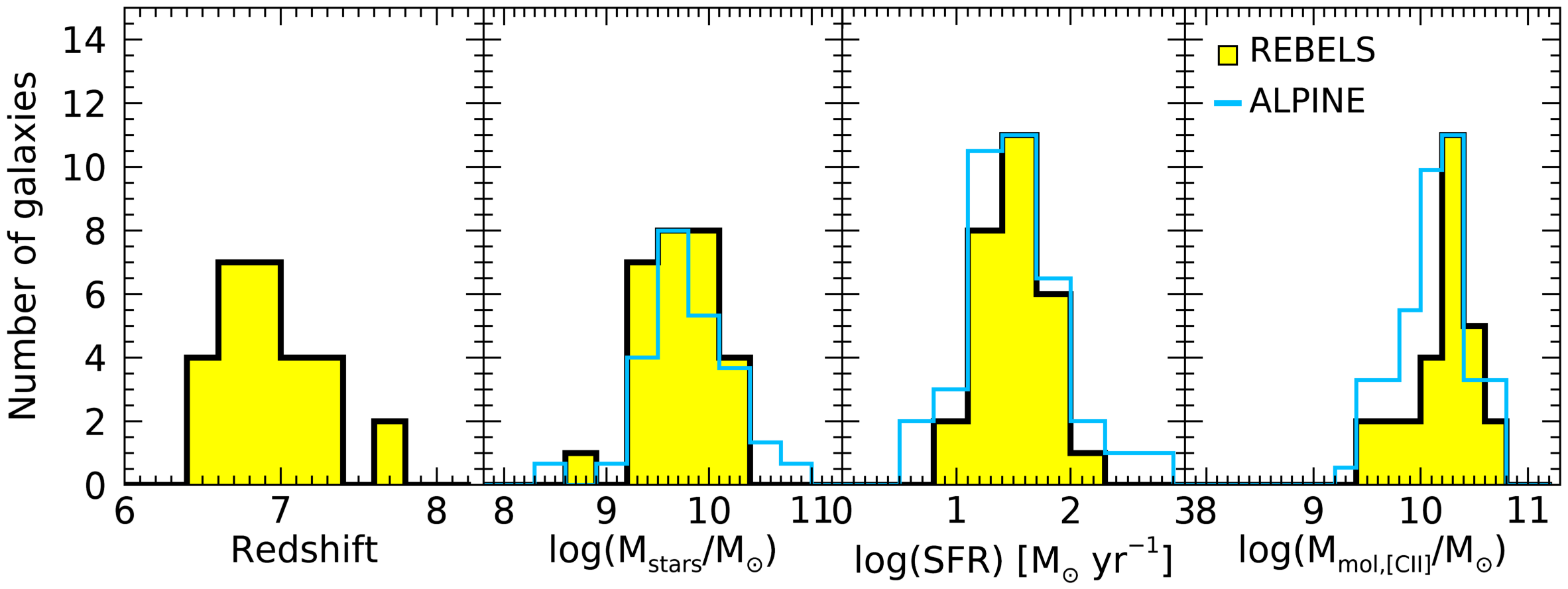}
    \caption{Distribution of properties of the REBELS galaxies (yellow filled histogram) compared to the ALPINE galaxies (blue empty histogram). From left to right, we show the distribution of redshift, stellar mass, SFR (UV+IR), and \Cii-based molecular gas masses. Stellar masses are derived from non-parametric SED fitting as described in the text \citep[see also,][]{topping22}. The histogram for the ALPINE sample has been normalized to the maximum number of REBELS galaxies. No comparison of the redshift distributions is provided, since, by construction, the samples cover different redshift ranges.  }
    \label{fig:distr}
\end{figure*}

Data calibration was done using the Common Astronomy Software Application (CASA) software following the observatory-provided pipeline. Datacubes were thus created, inverting the calibrated visibilities with the \texttt{tclean} task and using a natural weighting scheme to maximize the sensitivity. Continuum images were obtained by collapsing the datacubes and excluding the frequency ranges that contained the \Cii{} lines (when detected) within $2\times$FWHM$_{\rm [CII]}$. The \Cii{} line datacube and the continuum image were searched for emission at the location of the rest-frame UV source, following the procedures detailed in Schouws et al. (in preparation) and \citet{inami22}.

We obtained 23 \Cii{} line and 16 dust continuum detections, respectively, from the first year data. In this study, we thus focus on 23 galaxies with observations of \Cii{} line and dust continuum emission at $z\sim7$, plus additional five galaxies with \Cii{} detections (two of them with dust detections) from the pilot studies \citep{smit18, schouws22b}. We deliberately exclude [CII] non-detections from the analysis, as it is unclear whether they are [CII] faint galaxies or they are sources with redshifts falling outside the line scan covered ranges. The redshift distribution of the 28 confirmed \Cii{} line emitters considered in this work is shown in Figure \ref{fig:distr} (left). The mean redshift of these sources is $z=6.95$, spanning a range of $z=6.50-7.68$. 

\subsection{Ancillary data}
\label{sec:ancillary}
The REBELS sample was drawn from various cosmological survey fields, including the COSMOS/UltraVISTA \citep{scoville07, mccracken12}, UKIDSS/UDS and VIDEO/XMM-LSS \citep{jarvis13, lawrence07}, CANDELS \citep{grogin11}, CLASH \citep{postman12}, and BoRG/HIPPIES pure-parallel \citep{trenti11,yan11}. The multi-wavelength imaging of these surveys provided full optical to near-IR photometric coverage. 

The available photometry and the \Cii-derived spectroscopic redshifts were thus used to infer the galaxies' physical properties, including stellar masses as described by \citet{topping22}. Estimates of total SFRs was done through rest-frame UV observations from HST \citep{bouwens22} and estimates of the IR luminosities from the ALMA rest-frame $158\mu$m continuum measurements \citep[for details see:][]{inami22}.

Spectral Energy Distribution (SED) fitting for the REBELS sample was conducted using the \textsc{Prospector} code \citep{leja19}. Two different approaches were employed: a non-parametric prescription for the galaxies' star formation histories (SFHs) \citep[SFHs; ][]{johnson21} and a constant SFH. In both methods, consistent physical assumptions were made, including sub-solar metallicities ($0.2\ Z_\odot$), a \citet{chabrier03} initial mass function (IMF) for stellar masses below $300\ M_\odot$, and a Small Magellanic Cloud (SMC) dust extinction curve \citep[see:][]{topping22}.

Additionally, SED fitting for the REBELS sample was performed using the \textsc{Beagle} code \citep{chevallard16} with a constant SFH (Stefanon et al., in preparation) and identical physical assumptions, except for the fitting code and stellar population templates. Some of the derived properties obtained through this procedure have been reported in \citet{bouwens22}.

As elaborated in detail by \citet{topping22}, the use of \textsc{Prospector} with a constant SFH yields results similar to those obtained by \textsc{Beagle}. However, it is known that adopting constant SFHs can result in younger age estimates for starbursting galaxies, as the starburst can dominate the signal over older stellar populations (i.e. the `outshining' problem: young stellar populations outshine the old ones). In such cases, non-parametric SFHs provide greater flexibility and can help mitigate this issue \citep{algera23a}. Nevertheless, these models tend to yield larger estimates of stellar masses, with an average increase of 0.43 dex compared to constant SFH models \citep{topping22}. For consistency with a related parallel study of the HI mass content of REBELS galaxies \citep{heintz22}, we adopt the results from \textsc{Prospector} with non-parametric SFHs for this work. Using this approach, the majority of the REBELS galaxies are consistent within the uncertainties with the standard star formation main-sequence extrapolated to $z=7$ \citep{algera23a}. In Section \ref{sec:biases}, we discuss potential biases and the effect on our results introduced by the use of these non-parametric models.

\subsection{Comparison samples}
\label{sec:samples}

Our prime comparison sample is based on the ALPINE survey \citep{lefevre20}. This survey obtained \Cii{} emission line and dust continuum observations for a sample of 118 star-forming galaxies in the redshift range $z=4-6$ \citep{bethermin20}. These galaxies were selected based on their rest-frame UV properties and the availability of rest-UV spectroscopic redshift measurements. The derived stellar masses, in the range $10^{8.4}-10^{11}\ M_\odot$, and SFRs, in the range $3-630$ $M_\odot$ yr$^{-1}$, follow the expected trend of the star-forming main sequence at $z=4-6$. The ALPINE observations yielded [CII] line and dust continuum detections for 75 and 23 sources, respectively, at a signal-to-noise level of 3.5. While the range of stellar masses and SFRs covered by ALPINE is similar to that of REBELS (see Fig. \ref{fig:distr}), by design, the latter did not require the existence of previous rest-UV (e.g., Ly-$\alpha$) redshift determinations. 

\begin{figure}
    \centering
    \includegraphics[scale=0.56]{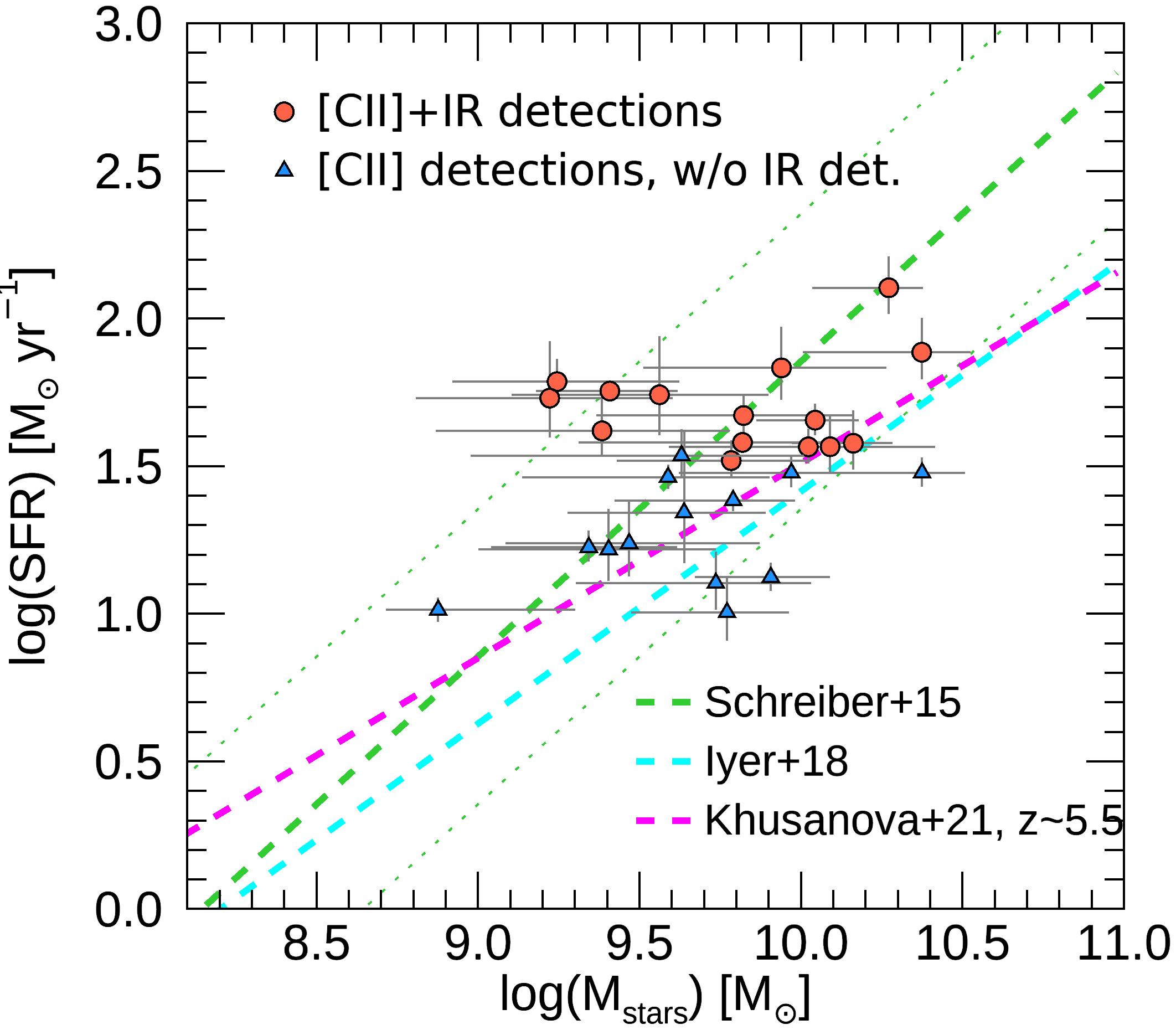}
    \caption{SFR(UV+IR) vs. stellar mass diagram for the REBELS sources. Blue triangles show sources detected in \Cii{} emission without IR continuum detection. Red color circles show sources detected in \Cii{} and IR emission. The dotted and dashed lines represent models of the star-forming main sequence at $z=7$ \citep{schreiber15, iyer18} and at $z\sim5.5$ \citep{khusanova21}. Most of the REBELS galaxies are consistent with being within $\pm0.3$ dex of the expected main sequence at $z=7$ \citep[see][]{topping22}.}
    \label{fig:ms}
\end{figure}

Figure \ref{fig:distr} (middle panels) shows the distribution in stellar mass and SFR for the REBELS galaxies, compared with that of the ALPINE sample. Our current REBELS sample covers wide ranges in both parameters, with $M_{\rm stars}\sim10^{8.5}-10^{10.5}$ $M_\sun$ and SFR$\sim30-300$ $M_\sun$ yr$^{-1}$, matching the range covered by ALPINE. Figure \ref{fig:ms} shows the location of the REBELS sources in the SFR(UV+IR)--stellar mass plane, compared to the expected location of the star-forming main sequence (MS) at $z\sim5-7$, based on various prescriptions \citep{schreiber15, iyer18, khusanova21}. For clarity, we split our sample into sources detected and not in 158$\mu$m dust continuum emission. Most galaxies in our sample of [CII] detected sources fall within the boundaries of the MS at these redshifts ($\pm0.3$ dex) compared to the widely used \citet{schreiber15} prescription for galaxies at $z=7$. However, they appear to be above the MS compared with the \citet{iyer18} prescription at $z=7$, although still within $\pm0.5$ dex. 

Based on this comparison, we find that REBELS galaxies have parameter distributions similar to the ones from the ALPINE sample. We note that the REBELS galaxies were not confirmed spectroscopically through rest-UV lines (Lyman-$\alpha)$ as was the ALPINE sample, although both samples were selected based on their rest-UV brightness. Furthermore, we find that most REBELS galaxies are consistent with being in the MS and thus represent examples of massive yet ``typical'' star-forming galaxies at high redshift. A similar conclusion is found by a recent study by \citet{dicesare23}.

At lower redshifts ($z<4$), we adopt the compilation of CO line and dust continuum measurements obtained by the IRAM Plateau de Bureau HIgh-z Blue Sequence Survey \citep[PHIBSS; ][]{tacconi13, tacconi18, freundlich19}. 

The PHIBSS is a targeted survey that provides stellar masses, SFRs, and molecular gas masses (from dust and CO observations) for large samples of 1444 massive star-forming galaxies at $z=0-4$.

We complement this with similar data from the ALMA Spectroscopic Survey in the Hubble UDF \citep[ASPECS; ][]{walter16, aravena19, aravena20, boogaard20} and other surveys \citep{papovich16, magdis17, gowardhan19, kaasinen19, molina19, bourne19, pavesi18, cassata20}. We remark that since the [CII] emission line is difficult to access at these lower redshifts, the molecular gas mass estimates in the comparison samples are based on observations of CO and dust continuum emission.  


\section{Measurements of molecular gas}
\label{sec:gasestimates}

Several far-infrared to radio emission line and dust continuum measurements have been used as standard tracers of the molecular gas content in galaxies \citep[e.g.,][]{carilli13, hodge20}. These include the CO line, particularly in the lowest energy transitions, the \Ci{} line, and dust continuum emission in the Rayleigh-Jeans regime. These tracers have provided molecular gas mass measurements in galaxies out to $z\sim3-4$. However, due to their faintness, these tracers become harder to detect at high redshifts with current facilities. Furthermore, at $z>6$, the Cosmic Microwave Background (CMB) temperature becomes comparable to the excitation temperature of the most standard tracers like CO $J<3$ and dust emission in the Rayleigh-Jeans tail \citep{dacunha13}. The radiation from these tracers thus becomes almost impossible to detect against the CMB at these redshifts.

The \Cii{} 158$\mu$m line, due to its brightness, high excitation temperature, and accessibility from the ground at the observed redshifted frequencies (due to the atmosphere transparency), has appeared as a top ISM tracer in galaxies at $z>4$. \citet{zanella18} found a correlation between the \Cii{} luminosity and the molecular gas mass with a mean absolute deviation of 0.3 dex, and without evident systematic variations. They find a [CII] luminosity to molecular gas mass conversion factor $\alpha_{\rm [CII]}\simeq 30 \ (M_\sun/L_\sun)$. As indicated by \citet{zanella18}, the \Cii-to-H2 conversion factor, $\alpha_{\rm [CII]}$, appears to be largely independent of the molecular gas depletion time, metallicity, and redshift for galaxies in the star-formation main sequence. Such correlation appears to hold for galaxies on and above the MS, in the redshift range $0<z<6$, and with metallicities 12+log(O/H)=7.9 to 8.8. A larger scatter is observed for galaxies above the MS at lower metallicities.

Similar alternative correlations have been proposed most recently, although with varied values for the \Cii-to-H2 conversion factor. \citet{madden20} find $\alpha_{\rm [CII]}\simeq 130 \ (M_\sun/L_\sun)$ based on local galaxy observations with mostly low metallicites, while lower values are obtained based on simulations for high redshift galaxies ($z\sim6$), $\alpha_{\rm [CII]}\simeq 10-18 \ (M_\sun/L_\sun)$ \citep[e.g., ]{pallotini17, vizgan22}. The discrepant $\alpha_{\rm [CII]}$ values in these studies seem to arise from competing unaccounted CO-dark gas and low metallicity effects, respectively.

There is vast literature indicating that the \Cii{} emission in galaxies arises from various ISM phases, including ionized, neutral, and molecular gas \citep[for a detailed discussion see: ][]{lagache18, ferrara19, pallotini19, dessauges20}. Numerical simulations and observations have shown evidence that $>60\%$ of the \Cii{} emission is dominated by molecular clouds and photon-dominated regions (PDRs), with a minor contribution of $20-25\%$ each from diffuse ionized and neutral gas \citep{vallini15, olsen17, katz17, katz19, pallotini17, pallotini22, vizgan22}. As such, \Cii{} can be readily calibrated as a tracer of these various phases, including star formation \citep[e.g.,][]{delooze14}, neutral gas \citep{heintz21} and molecular gas \citep{zanella18, madden20}. These calibrations are a likely function of stellar mass (and/or metallicities), evolutionary stage, and cosmic time. The main caveat to using \Cii{} as a molecular gas tracer is the unknown fraction of \Cii{} emission that arise from PDRs. However, the current results indicate that \Cii{} is a tracer as good as others, such as CO and dust for galaxies with stellar masses $\sim10^{10}\ M_\sun$, particularly considering the need to assume an excitation ladder and a metallicity-dependent $\alpha_{\rm CO}$ conversion factor for CO measurements, and given the necessary assumption of a gas-to-dust conversion (or similar calibrations) for dust continuum based measurements. 

Based on this, we measured molecular gas masses for the REBELS galaxies in our sample using Equation (2) from \citet{zanella18}, and following the approach adopted for the ALPINE survey by \citet{dessauges20},
\begin{equation}
{\rm log} (L_{\rm [CII]}) = (-1.28\pm0.21) + (0.98\pm0.02)  {\rm log}(M_{\rm{mol}})
\end{equation}

We thus adopt $\alpha_{\rm [CII]}\simeq30 (M_\sun/L_\sun)$, with $M_{\rm{mol}}=\alpha_{\rm [CII]} L_{\rm [CII]}$. Figure \ref{fig:distr} (right) shows the distribution of molecular gas masses obtained in this way. The masses range between $10^{9}-10^{11}\ M_\sun$, comparable to the range obtained for the ALPINE sample. Evidently, this is a consequence of the constant conversion factor and the same stellar mass and SFR ranges.

Before analyzing further the implications of these molecular gas measurements, in the rest of this section, we focus on checking the \Cii{}-based molecular gas mass estimates against other derivations provided by the available data. 

\begin{figure*}
    \centering
    \includegraphics[scale=0.42]{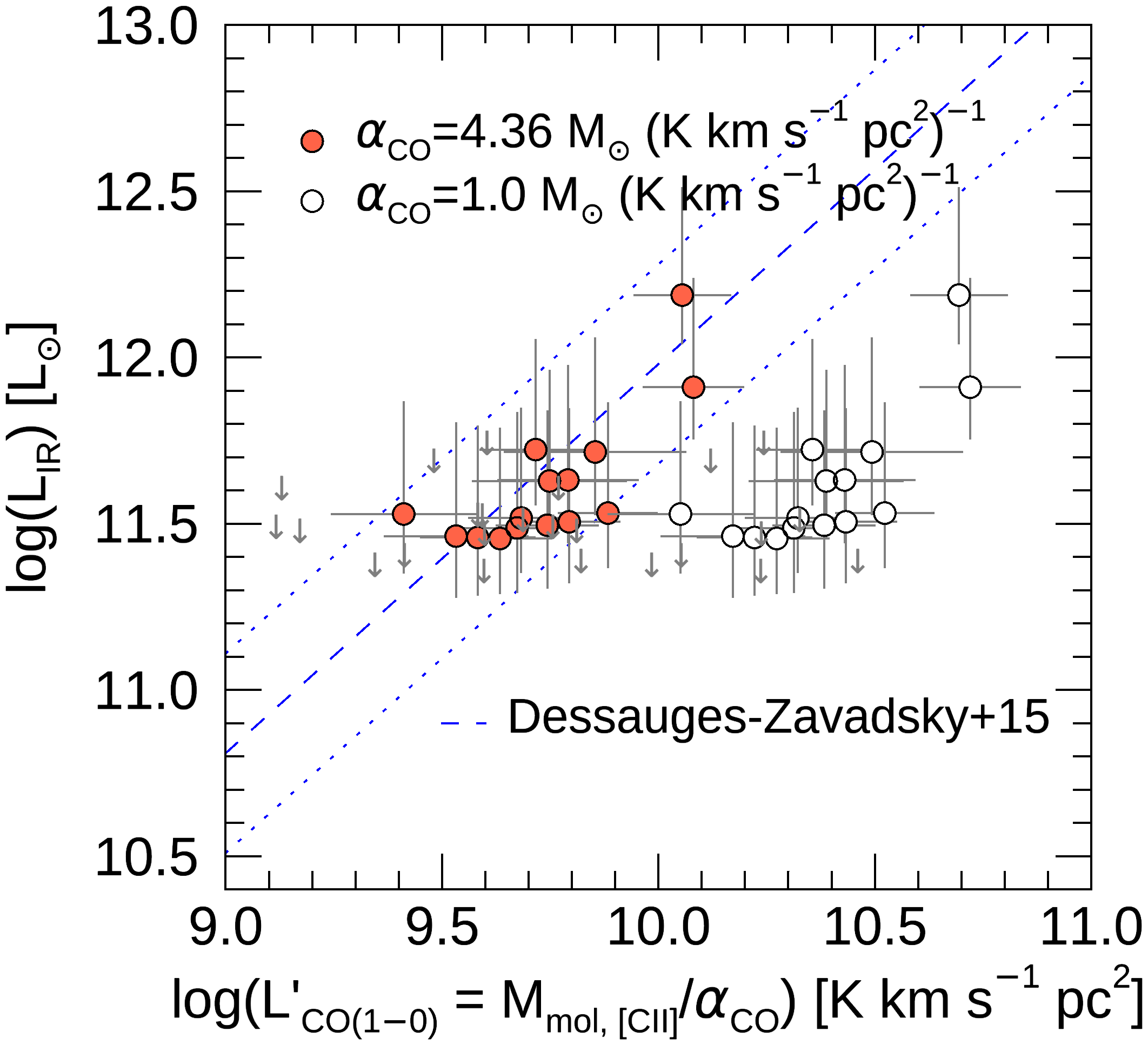}
    \includegraphics[scale=0.42]{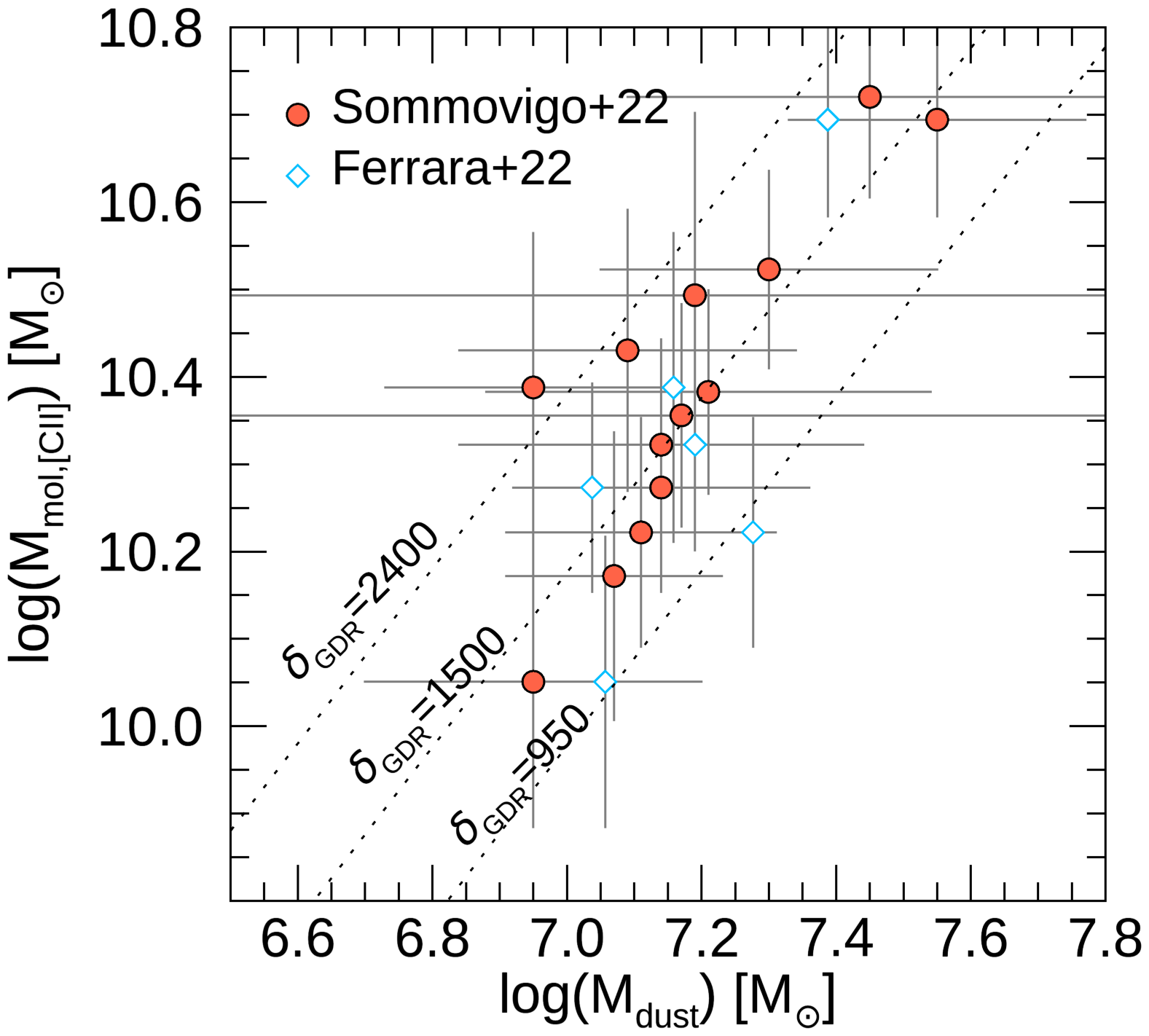}
    \includegraphics[scale=0.42]{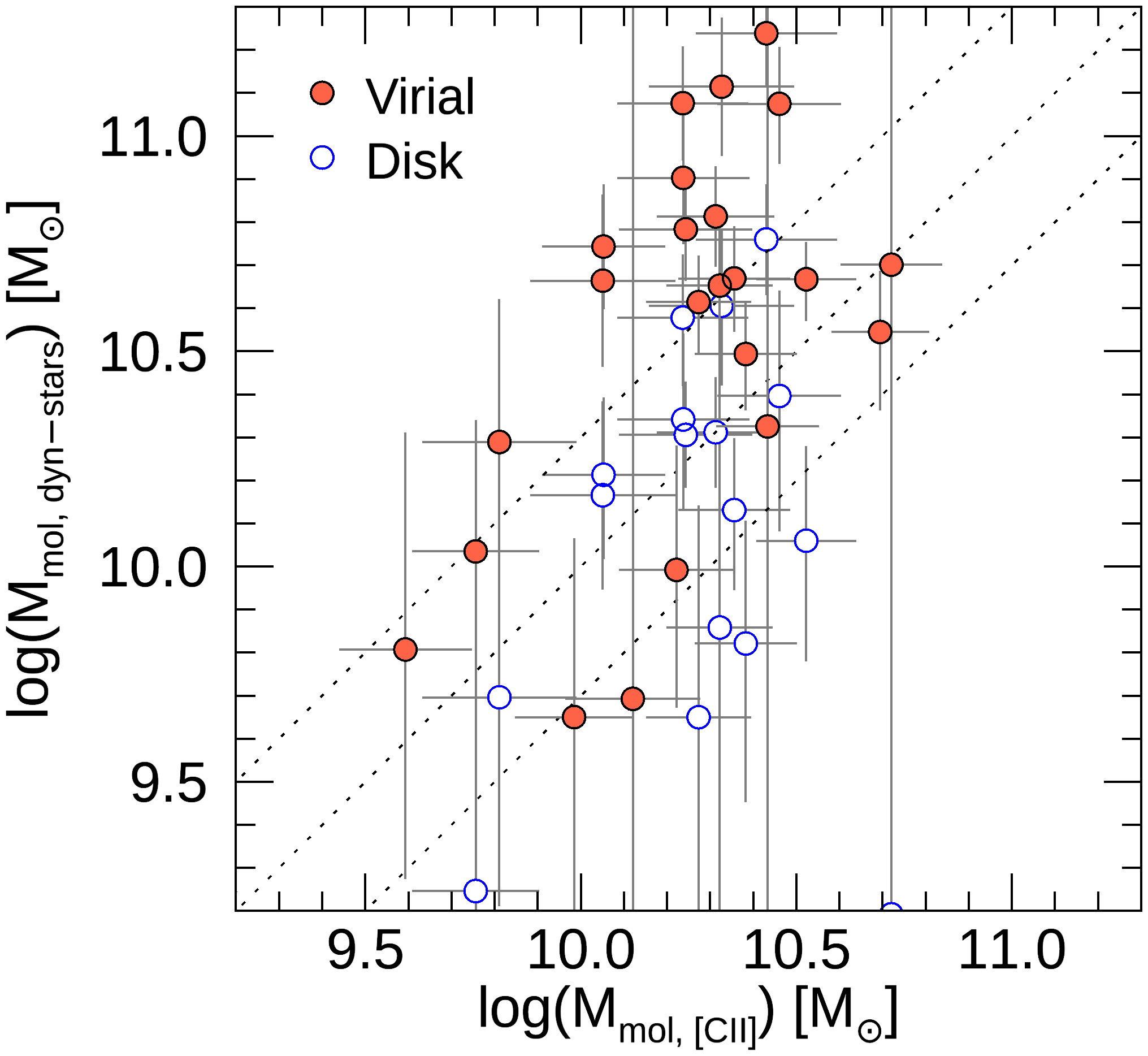}
    \caption{({\it Left:}) $L_{\rm IR}$ versus $L'_{\rm CO(1-0)}$ for the REBELS galaxies. The filled orange circles show the result of assuming a ``Milky Way'' conversion factor, $\alpha_{\rm CO}=4.6$, while the open circles show the case of $\alpha_{\rm CO}=1.0$. The linear fit to \aco{} observations of local spiral and disk galaxies at high redshift compiled by \citet{dessauges15} is shown as a dashed blue line \citep[see also,]{daddi10, aravena16a}. The dotted line shows the same line, with a factor of $\pm0.3$ dex, representing the typical scatter. ({\it Middle:}) Comparison of the \Cii-based molecular gas masses for the REBELS galaxies with the dust mass estimates from the models from \citet{ferrara22} and \citet{sommovigo22} are shown as blue diamonds and orange circles, respectively. Dotted lines highlight curves of constant gas-to-dust ratios. ({\it Right:}) Comparison of the \Cii-based molecular gas masses ($M_{\rm mol, [CII]}$) with estimates obtained from the dynamical and stellar masses ($M_{\rm mol, dyn-stars}$). Derivations of the dynamical mass based on an assumed spherical (virial) and disk geometries are shown as orange-filled and empty circles, respectively. The dotted lines represent the 1:1 relation and the range $\pm0.3$ dex. }
    \label{fig:checks}
\end{figure*}

\subsection{IR luminosity as a proxy of molecular gas mass}

One of the best-established relations in extragalactic astronomy is that between the IR luminosity ($L_{\rm IR}$) and the CO(1-0) luminosity ($L'_{\rm CO}$). This has long been known to hold for a variety of galaxies and environments, from local spirals \citep{leroy08} to the most extreme starburst galaxies known through a wide range of redshifts, out to $z\sim6$ \citep[e.g.,][]{aravena16a}.  Following \citet{dessauges20}, we convert the \Cii-derived molecular gas masses ($M_{\rm mol, [CII]}$) into the expected \aco{} luminosities under the assumption of typical $\alpha_{\rm CO}$ values, with $L'_{\rm CO}=M_{\rm mol, [CII]}/\alpha_{\rm CO}$. The $\alpha_{\rm CO}$ is known to depend on the metallicity, although with significant scatter \citep[e.g.,][]{bolatto13}, and thus a range of typical values is considered between 1.0 to 4.36 $M_\sun$ (K km s$^{-1}$)$^{-1}$. 

Figure \ref{fig:checks} (left) shows the $L_{\rm IR}$ vs. $L'_{\rm CO}$ diagram for the REBELS targets compared to standard curves representing the ranges occupied by star-forming galaxies, for reference. This includes the universal fit to the compilation of CO observations for star-forming galaxies by \citet{dessauges15}, along with the same line with a factor $\pm0.3$ dex added, representing the typical scatter around the fit. 

We find that assuming a standard value of $\alpha_{\rm CO}=4.36$ $M_\sun$ (K km s$^{-1}$)$^{-1}$, similar to what is found in the Milky Way and main-sequence galaxies at $z\sim2$, the REBELS galaxies fall right on top of the \citet{dessauges15} curve. Variations of $\alpha_{\rm CO}$ between $2.4-6.4$ $M_\sun$ (K km s$^{-1}$)$^{-1}$ would still yield a correlation within the 0.3 dex of the $L_{\rm IR}$ vs. $L'_{\rm CO}$ relation. Conversely, using a lower $\alpha_{\rm CO}$ of 1.0 (same units), comparable to what is found in local starbursts \citep{downes98}, would lead to significantly larger CO luminosities  being inconsistent with the $L_{\rm IR}$ vs. $L'_{\rm CO}$ relations. Using values of $\alpha_{\rm CO}>6.4$ $M_\sun$ (K km s$^{-1}$)$^{-1}$, as expected for sub-solar metallicities for high redshift galaxies, would also lead to $L'_{\rm CO}$ values incompatible with the $L_{\rm IR}$ vs. $L'_{\rm CO}$ relation. 

We note that the values of $L_{\rm IR}$ appear to have a threshold at $\sim10^{11.5} L_\sun$, which is due in part to the depth of the REBELS program. However, we caution that the REBELS $L_{\rm IR}$ values were computed using a single dust continuum measurements and realistic assumptions of the dust temperature (46 K). Warmer (or colder) dust temperatures will yield larger or lower $L_{\rm IR}$ values. 

While it is not surprising to find a linear scaling between both quantities, as the \Cii{} luminosities are known to scale with SFR, the fact that the \Cii-based CO values are consistent with the known  $L_{\rm IR}$-$L'_{\rm CO}$ relation supports the idea of using \Cii{} as a molecular gas tracer in the REBELS sample. 


\subsection{Dust masses}

The dust content in galaxies is related to the molecular gas mass through the gas-to-dust ratio ($\delta_{\rm GDR}$). This parameter depends strongly on the galaxies' metallicities and thus on the stellar mass through the mass-to-metallicity (MZ) relation. 

Dust mass ($M_{\rm d}$) estimates for the REBELS galaxies have recently been obtained by two independent studies by \citet{ferrara22} and  \citet{sommovigo22}. It is thus interesting to check how well the dust masses derived by these studies relate to the \Cii-based molecular gas measurements in the REBELS sample. The former constructed a semi-empirical model for dust reprocessing that uses the UV spectral slope, the observed UV continuum flux at 1500\AA \ and the observed FIR continuum at 158$\mu$m as input parameters, and considers various extinction curves and dust geometries. The latter adopted the \citet{sommovigo21} method to derive the dust properties of galaxies based on a combination of the \Cii{} line emission and the underlying 158$\mu$m continuum, using standard functional forms with varying normalization for the star formation law \citep[``Schmidt-Kennicutt'' relation][]{ferrara19, pallotini19} and standard assumptions for the dust properties \citep[][]{sommovigo21, sommovigo22}. 

Adopting a prescription of the MZ relation \citep[e.g.,][]{genzel15}\footnote{We adopt this prescription as it is widely used in similar studies at high redshift, as in \citet{tacconi18}.} and the broken power-law form of the $\delta_{\rm GDR}$-Z relation prescribed by \citet{remy14}, we find typical values of $\delta_{\rm GDR}=1000-3000$ for the stellar masses and redshifts obtained for the REBELS galaxies (median $\sim10^{9.7}\ M_\sun$, $z\sim7$). These $\delta_{\rm GDR}$ values are consistent with recent predictions for these stellar masses and redshifts, based on hydrodynamical simulations \citep[\texttt{DustyGadget}; see Fig. 4 from][]{graziani20}.

Figure \ref{fig:checks} (middle) shows the comparison of both $M_{\rm d}$ estimates and $M_{\rm mol, [CII]}$, in the available galaxies. We find that both dust mass estimates are consistent and linearly related to the \Cii-based molecular gas mass estimates within the scatter ($\sim30\%$), suggesting $\delta_{\rm GDR}\sim1500$. While it is not surprise that our molecular gas estimates correlate well with the dust mass estimates from \citet{sommovigo22}, as both use [CII] line information, the comparison suggest an overall consistency among the different methods and measurements, including the independent method used by \citet{ferrara22}.

\subsection{Estimates from dynamical mass}

An alternative way to compute the molecular gas mass is provided by considering the mass budget in each galaxy. The dynamical mass within half-light radius will be given by $M_{\rm dyn}(<r_{\rm e})=0.5(M_{\rm stars}+M_{\rm gas})+M_{\rm DM}(<r_{\rm e})$ \citep[e.g.,][]{daddi10}. Assuming that the mass budget within $r_{\rm e}$ will be dominated by baryons, thus the dark matter fraction is negligible and that the ISM is dominated by the molecular gas ($M_{\rm gas}\simeq M_{\rm mol}$), we can use the dynamical and stellar masses to obtain $M_{\rm mol, dyn}$. 

Unfortunately, our \Cii{} observations can spatially resolve only mildly some of the sources, and therefore we do not have information about their resolved \Cii{} kinematics. 

For simplicity, we estimate the dynamical masses of the REBELS sources using equations 10 and 11 in \citet{bothwell13}. In the first case, the gas is distributed in a virialized spherical system, e.g. a compact starburst, with $M_{\rm dyn}=1.56\times10^6 \sigma^2 R$, where $\sigma$ is the velocity dispersion ($\sigma=\Delta v_{\rm FWHM}/2.35$) and $R$ is the source radius. In the second case, the gas is distributed in a disk with $M_{\rm dyn}=4\times10^4 \Delta_{\rm FWHM}^2 R/sin(i)^2$, where $i$ is the inclination angle of the disk. In both cases, we estimate the source radius from the circularized estimate $R=\sqrt{a_{\rm maj}\times a_{\rm min}}$, where $a_{\rm maj}$ and $a_{\rm min}$ are the semi-major and semi-minor axes, respectively. We restrict the estimates on $R$ by considering only sources that have been resolved at least marginally in a given axis with a significance of 2, or $a/\delta a>2$. In unresolved cases, we consider the source to be small and simply fix their radius to 2 kpc, based on previous size measurements in galaxies at similar redshifts \citep{fujimoto20, herreracamus21} and for the REBELS sample \citep{fudamoto22}. This translates into $0.36''$ at $z=7$ (or a source diameter of $\sim0.7''$), corresponding to $\sim25\%$ of the observed beam size ($1.6''$). In the disk case, we adopt a mean inclination angle sin$(i)=\pi/4\approx0.79$. The virialized spherical geometry dynamical mass estimator will yield a mass 4.4 larger than the one obtained by assuming a disk-like gas distribution for the same source size, linewidth, and an average inclination parameter.  

Figure \ref{fig:checks} (right) shows the comparison between the \Cii-based molecular gas mass ($M_{\rm mol, [CII]}$) and the one estimated from the dynamical and stellar masses ($M_{\rm mol, dyn-stars}$). From Fig. \ref{fig:distr} we note that the stellar masses are systematically smaller than the derived molecular gas masses and thus have little effect in this comparison. Despite the number of assumptions and uncertainties in the parameters, we find overall agreement between both molecular gas estimates in the case of the disk geometry, with most of the galaxies falling within $\pm0.3$ dex from a linear relation. In the case of virialized geometry, we find a systematic offset between the \Cii{} \ and dynamical estimates and large scatter. The reason for such large scatter is likely due to the simple geometry assumption for very complex star-forming systems at these redshifts, where most of them are not expected necessarily to be either rotating disks or fully spherical systems. 


In summary, we find that all the independent checks obtained by comparing the \Cii-based molecular gas mass estimates with the IR luminosity, the dust masses, and the dynamical masses are consistent and support the application of the \citet{zanella18} calibration to estimate the molecular gas mass, assuming that the molecular gas-phase dominates the baryon budget at $z>6$ \citep[which might not be the case;][]{heintz22, vizgan22}. 

\section{Analysis and discussion}
\label{sec:analysis}

\begin{figure*}
    \centering
    \includegraphics[scale=0.62]{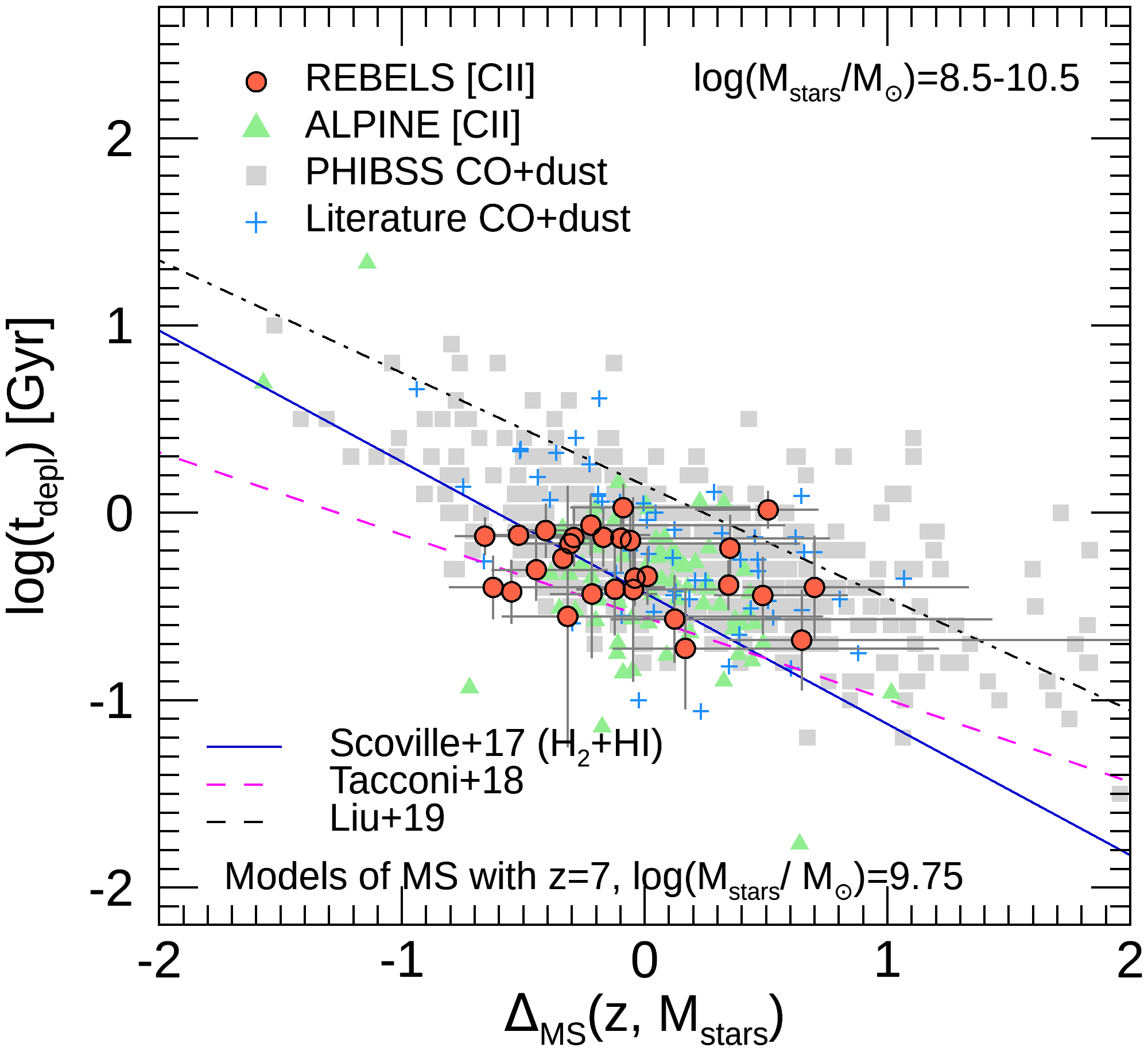}
    \includegraphics[scale=0.62]{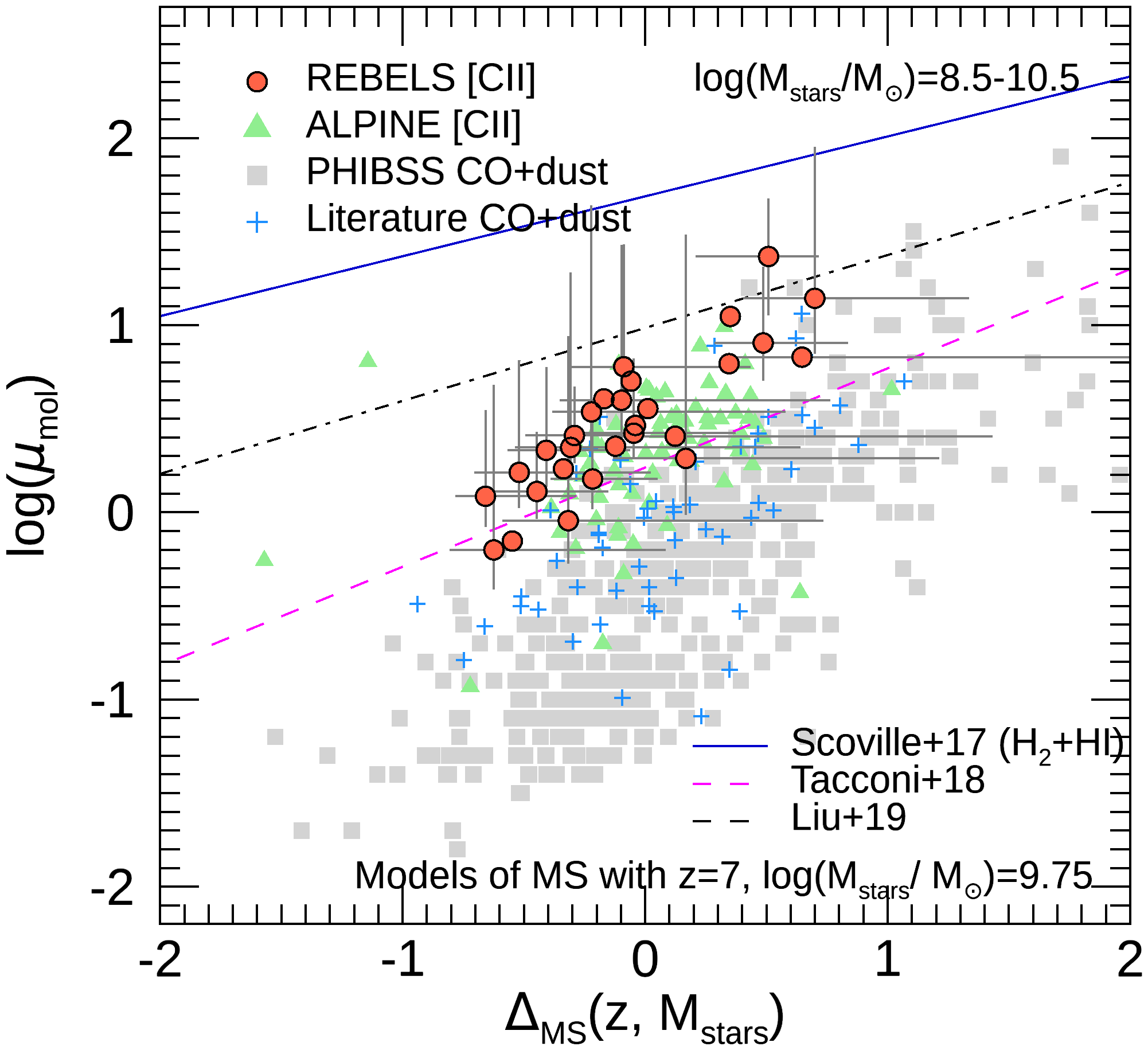}
    
    \caption{ISM scaling relations for the REBELS galaxies compared to literature.  ({\it Left:}) Molecular gas depletion timescale ($t_{\rm dep}$)  as a function of offset to the MS ($\Delta_{\rm MS}$). ({\it Right:}) Molecular gas to stellar mass ratio ($\mu_{\rm mol}$)  as a function of offset to the MS ($\Delta_{\rm MS}$) . In both panels, orange-filled circles show the REBELS galaxies $z=6-8$ with molecular gas masses derived from [CII] measurements. Green triangles show \Cii{} measurements in ALPINE galaxies at $z=4-6$ \citep{bethermin20}. Gray squares and blue crosses show CO and dust measurements from PHIBSS at $z<3$ \citep{tacconi18} and additional sources from the literature (see text), respectively. All samples have been restricted to galaxies with $M_{\rm stars}=10^{8.5-10.5} M_\sun$ to match the range spanned by REBELS galaxies. The solid blue, dashed magenta and black dot-dashed lines represent the prescriptions for the scaling relations computed for the median mass of the REBELS sample, $M_{\rm stars}=10^{9.75} M_\sun$ and $z=7$, from \citet{scoville17}, \citet{tacconi18} and \citet{liu19}, respectively.}
    \label{fig:scaling}
\end{figure*}

The relationship between the SFR and $M_{\rm mol}$ in galaxies is usually termed the global ``Schmidt-Kennicutt'' (SK) relation or ``star formation'' law. This relationship probes how star formation activity is linked to the molecular gas supply.  The ratio between the two quantities is usually called the molecular gas depletion timescale, $t_{\rm dep}=M_{\rm mol}/$SFR, which is seen as the amount of time left before the galaxy runs out of molecular gas at the current rate of star formation. Per definition, this quantity does not account for possible feedback processes in the galaxy (outflows) and interaction with the IGM (inflows). The fraction of baryonic mass in the galaxy that is in the form of molecular gas is usually described by the molecular gas fraction, $f_{\rm mol}=M_{\rm mol}/(M_{\rm mol}+M_{\rm stars})$ or as the molecular gas to stellar mass ratio, $\mu_{\rm mol}=M_{\rm mol}/M_{\rm stars}$.  

Several studies in the last decade have found that these quantities are intrinsically linked to other galaxy properties such as their stellar mass, specific SFR (sSFR), and/or their distance to the star-forming main sequence at a given redshift, $\Delta_\mathrm{MS} = \mathrm{sSFR}/\mathrm{sSFR}_\mathrm{MS}(z)$, yielding the differentiation of different modes of star-formation as a function of $\Delta_{\rm MS}$ \citep[e.g.,][]{sargent14}. Furthermore, $t_{\rm dep}$ and $\mu_{\rm mol}$ were found to evolve with redshift (and stellar mass). While mild evolution is observed for the average $t_{\rm dep}$ for galaxies at $z<3$, strong evolution is seen for the average $\mu_{\rm mol}$ \citep[e.g.,]{tacconi18, liu19}. Recent determinations of the molecular gas mass in galaxies from the ALPINE survey permitted extending these results to the redshift range $4-6$, indicating little evolution on the average values of both $t_{\rm dep}$ and $\mu_{\rm mol}$ from $z\sim4-6$ to $z\sim3$ \citep{dessauges20}, being therefore interesting to check if similar results are found in an independent sample, and at higher redshift.

In the following, we analyze and put in context the ISM properties of the REBELS galaxies compared to previous observations of distant galaxies. The comparison samples used are briefly described in Section \ref{sec:samples} and in the following analysis were restricted to the stellar mass range of the REBELS sample. This filter is important for the lower redshift samples which contain a significant number of more massive galaxies. 

\subsection{Scaling relations}

\begin{figure*}
    \centering
    \includegraphics[scale=0.62]{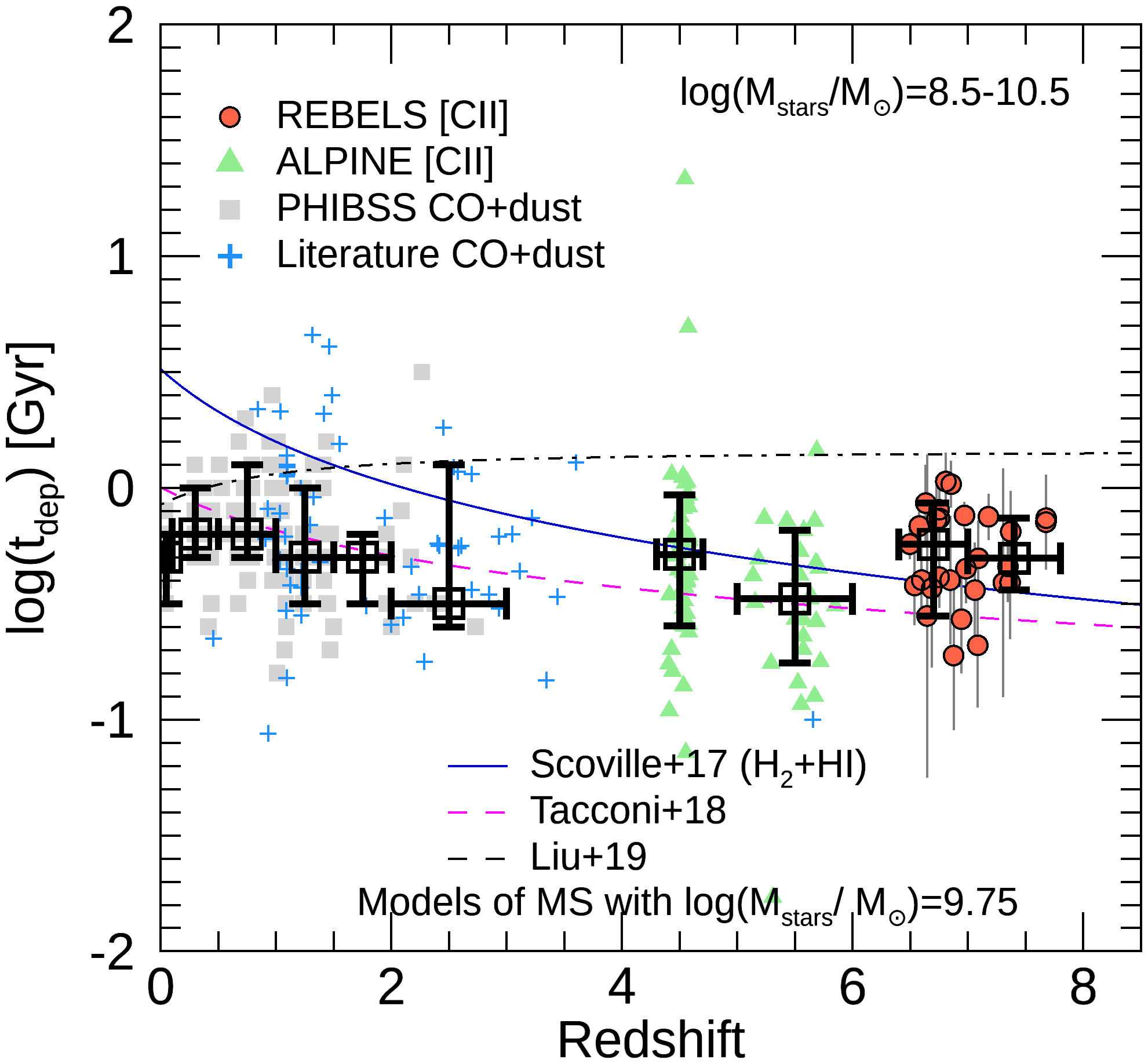}\hspace{1cm}
    \includegraphics[scale=0.62]{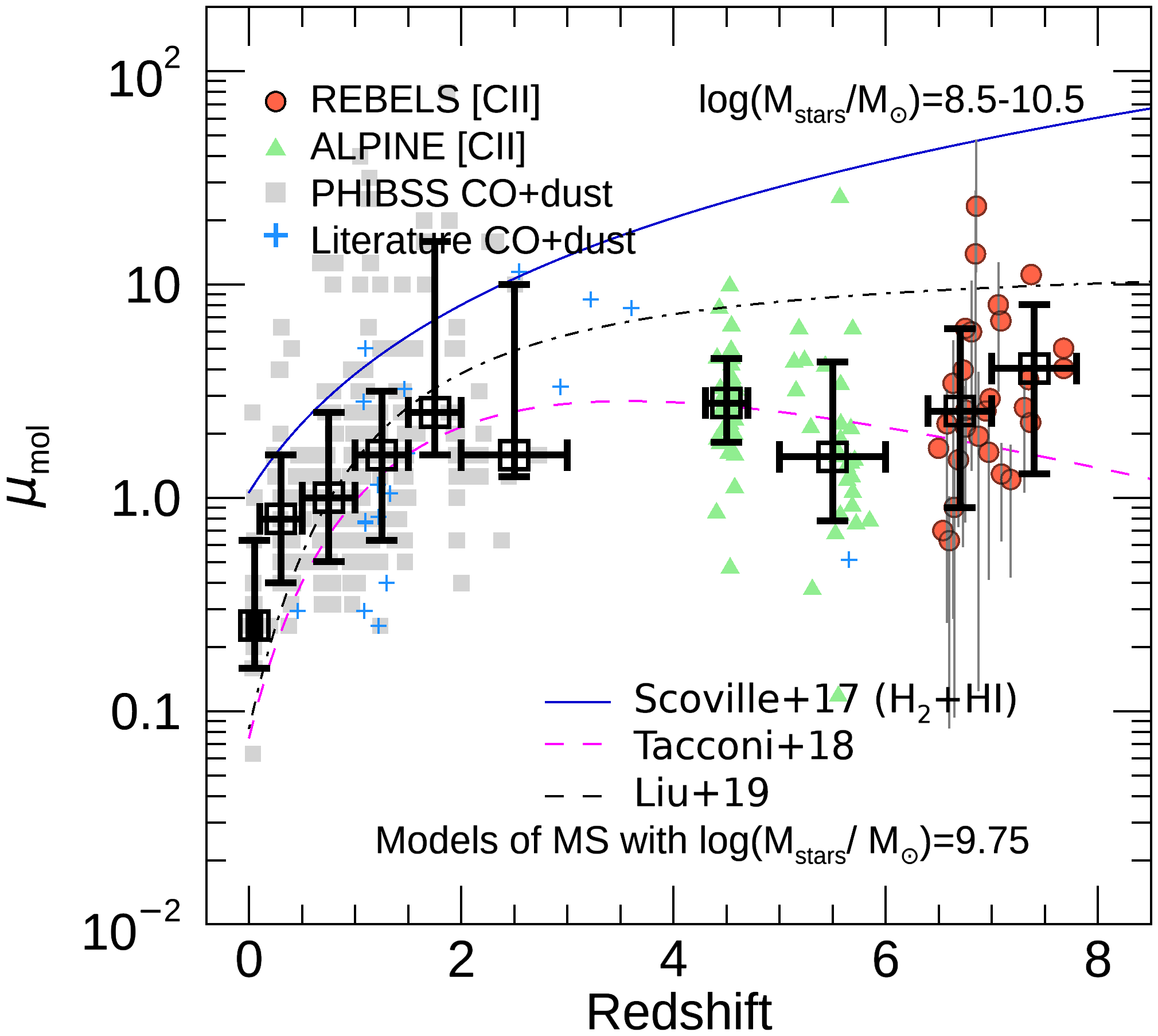}
    \caption{Evolution of the molecular ISM out to $z=7.5$. ({\it Left:}) Molecular gas depletion timescale ($t_{\rm dep}$) as a function of redshift. ({\it Right:}) Molecular gas to stellar mass ratio ($\mu_{\rm mol}$) as a function of redshift. Symbols are the same as in Fig. \ref{fig:scaling}. Similarly, literature comparison samples are trimmed to have the stellar mass range of the REBELS galaxies, $M_{\rm stars}=10^{8.5-10.5}\ M_\odot$. The black open squares represent the averages in each relevant redshift range.}
    \label{fig:zevol}
\end{figure*}

Figure \ref{fig:scaling} shows the relationships formed by $t_{\rm dep}$ and $\mu_{mol}$ with $\Delta_{\rm MS}$ for the REBELS sources, compared to the samples and prescriptions in the literature for the expected scaling relations at $z=7$ \citep{tacconi18, liu19, scoville17}, computed at the median stellar mass of the REBELS sample, $10^{9.5} M_\sun$ at $z=7$. We note that the \citet{scoville17} measurements provide the total ISM masses (HI and H$_2$), and thus, by construction, yield larger masses compared to molecular gas fractions only. For the MS, we use the prescription from \citet{schreiber15}. We note that these prescriptions for $\mu_{mol}$, $\Delta_{\rm MS}$ and the MS correspond to extrapolations based largely on CO and dust observations of galaxy samples $z<4$, as no similar observations are available for galaxies at $z>6$. 

The REBELS sample does not show a strong dependency of $t_{\rm dep}$ on $\Delta_{\rm MS}$ as opposed to lower redshift samples and expectations from scaling relations. This could be partly due to IR luminosity limits (see Fig. \ref{fig:checks}).

We find that all galaxies are consistent within the scatter with the expected low values for $t_{\rm dep}$ at these redshifts. A similar trend is followed by the ALPINE sample, indicating that this trend is not associated with sample selection.  Thus, the lack of dependency on $\Delta_{\rm MS}$ is likely related to the fact that most sources fall in the MS ($\Delta_{\rm MS}\sim0$) and to the scatter in this relationship expected at high-redshift. Moreover, the location of both REBELS and ALPINE samples in the $t_{\rm dep}$ vs. $\Delta_{\rm MS}$ plane agrees well with the expectation from the \citet{tacconi18} and \citet{scoville17} relations at these redshifts.

A significant difference in the dependency of $\mu_{mol}$ with $\Delta_{\rm MS}$ is seen between the high and low redshift samples and among the various prescriptions for the scaling relations. We find that for the REBELS and ALPINE samples, $\mu_{mol}$ consistently increases with distance from the MS, as expected (i.e. galaxies that have greater SFR per unit stellar mass have larger $M_{\rm mol}$ per unit stellar mass). Furthermore, $\mu_{mol}$ values are systematically larger for the REBELS and ALPINE samples compared to galaxies at $z\sim1-3$ (i.e. galaxies at higher redshifts are richer in molecular gas) for a given stellar mass. However, a significant difference is observed when comparing the various extrapolations for scaling relations at these redshifts, being the \citet{tacconi18} prescription the one that predicts values closer to our measurements. This is similar to previous findings from \citet{dessauges20}.

Overall, the REBELS (and ALPINE) galaxies mimic the scaling relation trends expected based on lower redshift samples. However, we find large discrepancies among the various models at these redshifts. This is particularly evident for the dependence of $\mu_{mol}$ on $\Delta_{\rm MS}$, where model predictions differ by nearly two orders of magnitude for MS galaxies. As mentioned earlier, an important source of uncertainty, which affects the values of $\Delta_{\rm MS}$ and $\mu_{mol}$, is the lack of sensitive constraints on the rest-frame optical SED. This will affect the stellar mass estimates and their derived uncertainties.


\subsection{Evolution of $t_{\rm dep}$ and $\mu_{\rm mol}$}

Figure \ref{fig:zevol} shows the measured dependency of $t_{\rm dep}$ and $\mu_{\rm mol}$ with redshift for the REBELS sources, compared with the literature samples and prescriptions for the expected scaling relations \citep{tacconi18, liu19, scoville17} for MS galaxies with a stellar mass of $10^{9.75} M_\sun$. To check for evolutionary trends, we divided the sample into ranges of redshift, grouping the ALPINE and REBELS samples into two bins each ($4.3<z<4.7$ and $5.0<z<6.0$ for ALPINE; $6.4<z<7.0$ and $7.0<z<7.8$ for REBELS) and dividing the literature galaxy samples at lower redshifts into six bins. With this, we computed the median among all galaxies in each bin. Uncertainties in each median point taken are obtained by computing the quartile values for the subsamples in each bin. These values are found to be in excellent agreement with weighted average values for the subsamples in each bin, and with previously reported values for these mass ranges \citep[e.g.,]{tacconi18, dessauges20}.



We find that the REBELS galaxies exhibit a mild (or no) evolution of $t_{\rm dep}$ from $z=7$ to $z\sim4-6$ (and at most redshifts), as expected from previous observations and models, yielding a median $t_{\rm dep}=0.5_{-0.14}^{+0.26}$ Gyr at $z=7$, where the uncertainty corresponds to the interquartile range. Little scatter is seen among the REBELS sample, and there is an overall consistency with the \citet{tacconi18} and \citet{scoville17} prescriptions. The short $t_{\rm dep}$ values found here require that there is additional gas accretion for the next 3 Gyr to sustain the high star formation activity until $z\sim2$.

Similarly, we find a median $\mu_{\rm mol}=2.6_{-1.4}^{+4.1}$ at $z=7$, equivalent to a molecular gas fraction $f_{\rm mol}=0.73_{-0.11}^{+0.17}$.   The scatter seen here is mostly due to the wide range of stellar masses of galaxies per redshift bin, since there is a strong dependence of $\mu_{\rm mol}$ on $M_{\rm stars}$ \citep[see Fig. 8 bottom panel from][]{dessauges20}. This scatter could also be caused by the fact that the REBELS galaxies are likely younger than lower redshift objects at matched stellar mass and thus caught at earlier stages in their assembly process, or simply because of the uncertainties in stellar mass measurements, given the lack of rest-frame optical photometry. 



In this case, the scaling relation model prescriptions diverge by one order of magnitude, with the \citet{tacconi18} model being the one that agrees best with the data. Since the \citet{scoville17} measurements yield the total ISM masses (HI and H2), they yield larger gas fractions when compared to molecular gas fractions only. 

\subsubsection{Potential effects of our adoption of stellar masses}
\label{sec:biases}
As discussed in Section \ref{sec:ancillary}, we have adopted stellar masses obtained through SED fitting using non-parametric SFHs as this approach helps overcome the outshining problem and for consistency with a parallel study \citep{heintz22}. For the sample considered in this study, these non-parametric stellar mass estimates are on average $\sim$0.35 dex higher than those computed using parametric constant SFHs. 

This selection has only a minor effect on the main conclusions drawn in this and in the previous section. The choice of non-parametric stellar masses make the REBELS galaxies mostly consistent with the star-forming main-sequence, whereas using constant SFHs yields a slightly higher fraction of starburst galaxies, or above 0.5 dex from the expected star-forming main sequence at $z\sim7$. We find that only two sources are above this threshold for non-parametric stellar masses versus 9 sources for parametric ones. 

The choice of non-parametric stellar masses has little effect on the computed $\Delta_{\rm MS}$. The median $\Delta_{\rm MS}$ for non-parametric SFHs is only 0.1 dex lower than that computed with parametric constant SFHs. Similarly, our choice of stellar mass measurements does not affect the $t_{\rm dep}$ estimates; however, it has a more important effect on $\mu_{\rm mol}$. The median $\mu_{\rm mol}$ obtained for non-parametric stellar masses is a factor 2.4 lower than that obtained with parametric ones. Choosing the parametric stellar masses would still put the REBELS galaxies within the expected scaling relations in the log($\mu_{\rm mol}$) vs. $\Delta_{\rm MS}$ plot (Fig. \ref{fig:scaling}). However, they would now be more consistent with the \citet{liu19} prescription for $\mu_{\rm mol}$ at $z=7$. This would be in contradiction to the consistently low value of $t_{\rm dep}$ at $z\sim7$, which is consistent with the prescription of \citet{tacconi20}. 

\subsection{Cosmic density of molecular gas}

Great observational efforts have been made in recent years to measure the CO luminosity function (LF) at different redshifts and, thereby, the cosmic density of molecular gas $\rho_{\rm mol}$. Initial attempts to measure the CO LF at high redshift were done with the Very Large Array (VLA) in an overdense field at $z\sim1.5$ \citep{aravena12}. However, the first dedicated survey for this purpose used the IRAM Plateau de Bureau Interferometer (PdBI) to perform a spectral scan over frequency and thus conduct a blind search for CO line emission in the {\it Hubble} Deep Field North \citep[$<1$ arcmin$^2$; ][]{walter14, decarli14}. Later on, CO line and dust continuum surveys with ALMA were performed to this end in the {\it Hubble} UDF, including the ASPECS pilot \citep[$\sim1$ arcmin$^2$,][]{walter16, aravena16b, aravena16c, decarli16a} and large programs \citep[$\sim5$ arcmin$^2$,][]{decarli19, gonzalezlopez20, aravena19, aravena20, decarli20, boogaard20}. Similar efforts were done with the VLA through the CO Luminosity Density at High-Redshift (COLDz) survey \citep{riechers19} covering larger areas at shallower depths for lower-J CO transitions. A parallel approach has taken advantage of targeted CO and dust continuum surveys, as each independently observed pointing would essentially yield a portion of a cosmological deep field, and combining it with statistical methods to assess the completeness and compute the cosmological volumes covered. This yielded estimates for the CO LF from the PHIBSS \citep{lenkic20}, A3COSMOS \citep{liu19}, and the ALMACAL surveys \citep{klitsch19}. Similar estimates have been obtained from dust continuum observations \citep{scoville17, magnelli20}, and most recently using the ALPINE \Cii{} survey \citep{yan20}.

\begin{figure*}

    \centering
    \includegraphics[scale=0.8]{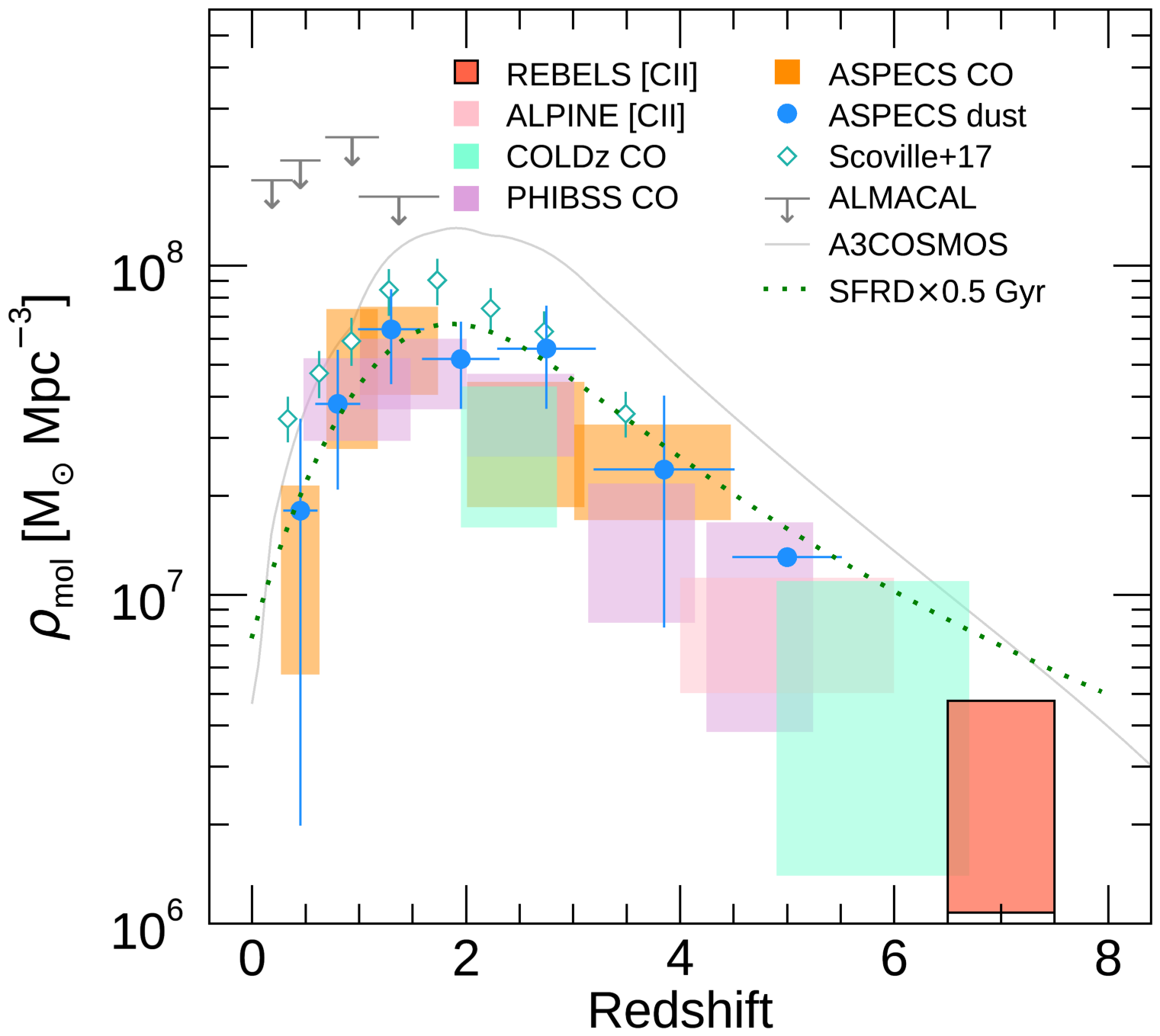}\hspace{1cm}
    \caption{Cosmic density of molecular gas ($\rho_{\rm mol}$) from REBELS at$z=6.5-7.5$, compared to measurements from the literature. The red-filled box region represents the \Cii-based measurement from this work. Literature data points come from various CO, dust, and \Cii{} surveys, including ASPECS CO \citep{decarli20} and dust measurements \citep{magnelli20}, COLDz \citep{riechers19}, PHIBSS \citep{lenkic20}, ALMACAL \citep{klitsch19}, A3COSMOS \citep{liu19}, dust measurements from \citet{scoville17} and ALPINE \citep{yan20}. The green dotted curve shows the cosmic SFR density \citep{madau14}, scaled by a typical molecular gas depletion timescale of 0.5 Gyr \citep{tacconi20}. REBELS measurements extending out to $z\sim7$ indicate a sustained decline of $\rho_{\rm mol}$ at earlier epochs}
    \label{fig:rhoh2}
\end{figure*}

It is interesting to compare these measurements with the value of $\rho_{\rm mol}$ at $z=6.5-7.5$ from the REBELS survey. Following previous studies \citep{yan20, heintz21, heintz22}, we compute $\rho_{\rm mol}$ from the \Cii{} luminosity density, $\mathcal{L}$, obtained for galaxies at $z\approx7$ from the REBELS survey by Oesch et al. (in preparation). In this study, the \Cii{} luminosity density is found by integrating the \Cii{} luminosity function down to a limit of log($L_{\rm [CII], lim})=7.5$, as $\mathcal{L}={\int_{\rm [CII], lim}^{\infty} L_{\rm [CII]} \phi(L_{\rm [CII]}) \,dL}$, where $\phi(L_{\rm [CII]})$ is the \Cii{} LF . The latter is obtained from the UV LF \citep{bouwens21}, using an empirical $L_{\rm [CII]}-L_{\rm UV}$ relation, derived for galaxies at the same redshift \citep[for details, see Oesch et al., in prep.; and also][]{barrufet23}.  With this, we obtain a  \Cii{} luminosity density log($\mathcal{L}/[L_\sun {\rm Mpc}^{-3}])=4.85^{+0.25}_{-0.20}$ for the REBELS sample. Now, this can be converted to a molecular gas mass density using the calibration described above, $\alpha_{\rm [CII]}\approx30 \, M_\sun \, L_\sun^{-1}$ \citep[from][]{zanella18}, yielding log($\rho_{\rm mol}/({\rm M}_\sun \, {\rm Mpc}^{-3}))=6.34_{-0.31}^{+0.34}$ at $z=6.5-7.5$. This result includes the uncertainties associated with the $\alpha_{\rm [CII]}$ calibration \citep{zanella18}. For the ALPINE data, we use the updated calculations of the \Cii{} luminosity density from \citet{heintz22}.

Figure \ref{fig:rhoh2} shows the result of this comparison. The measurements obtained from the different methods and tracers agree that $\rho_{\rm mol}$ resembles the evolution of the cosmic SFR density \citep[scaled by a typical molecular gas depletion timescale of 0.5 Gyr, shown by the solid green line in Fig. \ref{fig:rhoh2}; e.g.,][]{decarli20}, although the studies by A3COSMOS \citep{liu19} and \citet{scoville17} find different normalizations. 

However, the few measurements obtained at $z\sim4-6$, including the PHIBSS \citep{lenkic20} and COLDz \citep{riechers19} CO-based and the ALPINE \Cii{} based estimates \citep{yan20, heintz22}, appear to agree, indicating a continuous decline of $\rho_{\rm mol}$ at increasing redshifts. The REBELS measurements agree with the trend of $\rho_{\rm mol}$ declining at higher redshifts. These results indicate that $\rho_{\rm mol}$ increases by an order of magnitude from $z\sim7$ to $z\sim4$, and that beyond $z=6$, the values of $\rho_{\rm mol}$ reach levels similar and below the current $\rho_{\rm mol}$ (at $z=0$) of log($\rho_{\rm mol}/({\rm M}_\sun \, {\rm Mpc}^{-3}))\sim 7$.

\section{Conclusions}
\label{sec:conclusions}
We present molecular gas mass estimates and properties for the sample of galaxies at $z\sim7$ unveiled by the REBELS large program survey, using the \Cii{} line emission as a proxy for the molecular ISM. The main conclusions of this study are as follows.

\begin{enumerate}
\item Comparison of the \Cii-based molecular gas mass estimates with independent estimates and proxies, including the IR luminosity (as a proxy for the molecular gas mass through the $L_{\rm IR}-L'_{\rm CO}$ relation), dust masses, and dynamical mass estimates yield reasonable agreement when applied to the REBELS sample at $z\sim7$. Furthermore, the matched stellar mass and SFR distributions of REBELS and ALPINE surveys suggest that the calibrations to obtain $M_{\rm mol}$ from $L_{\rm [CII]}$ apply for massive main-sequence galaxies at $z\sim7$ and $z\sim4-6$.

\item The molecular to stellar mass ratios ($\mu_{\rm mol}$) and molecular gas depletion timescales ($t_{\rm dep}$) appear to follow the standard scaling relations with distance to the main sequence $\Delta_{\rm MS}$ for REBELS galaxies at $z\sim7$. While $\mu_{\rm mol}$ increases toward larger $\Delta_{\rm MS}$, $t_{\rm dep}$ is observed to stay relatively constant. The location of the REBELS galaxies at $z\sim7$ in these plots agrees with the regions occupied by ALPINE galaxies and the distant end of lower redshift samples at matched stellar masses.

\item The median $t_{\rm dep}$ and $\mu_{\rm mol}$ stay almost constant in the range $z\sim2-7$ for galaxies with stellar masses $10^{8.5-10.5}\ M_\sun$. 

\item We measure a value of the cosmic density of molecular gas at $z=6.5-7.5$ of log($\rho_{\rm mol}/({\rm M}_\sun \, {\rm Mpc}^{-3}))=6.34_{-0.31}^{+0.34}$, indicating an increase of an order of magnitude from $z\sim7$ to $z\sim4$.

\end{enumerate}

The advent of infrared facilities such as the JWST and Roman will allow us to explore samples of hundreds to thousands of galaxies at $z>6$, opening the study of the formation processes of galaxies through the end of the epoch of reionization. Investigating the cold ISM of such early galaxies will require observations with submillimeter facilities as ALMA. However, due to the dimming of traditional ISM tracers such as CO line and dust continuum emission due to cosmological distances, lower metallicities, and increasingly low-contrast to the cosmic microwave background at higher redshift \citep{dacunha13}, measurements of the ISM gas content (a key ingredient to study the galaxy assembly) will be increasingly difficult. The \Cii{} line emission has therefore come to save the day, enabling relatively accurate measurements of $M_{\rm mol}$ for massive systems in the early universe. Calibration of the \Cii{} line for larger samples of galaxies, as well as other fine-structure lines, will be essential for future galaxy formation studies.

\begin{acknowledgements}
We thank the anonymous referee for their constructive feedback that helped improve this paper. We thank Roberto Decarli for the useful discussions and for providing the cosmic density literature measurements. This paper makes use of the following ALMA data: ADS/JAO.ALMA\#2019.1.01634.L. ALMA is a partnership of ESO (representing its member states), NSF (USA) and NINS (Japan), together with NRC (Canada), MOST and ASIAA (Taiwan), and KASI (Republic of Korea), in cooperation with the Republic of Chile. The Joint ALMA Observatory is operated by ESO, AUI/NRAO and NAOJ. M.A. acknowledges support from FONDECYT grant 1211951, ANID+PCI+INSTITUTO MAX PLANCK DE ASTRONOMIA MPG 190030, ANID+PCI+REDES 190194 and ANID BASAL project FB210003. HI and HSBA acknowledge support from the NAOJ ALMA Scientific Research Grant Code 2021-19A.PD acknowledges support from the NWO grant 016.VIDI.189.162 (``ODIN") and from the European Commission's and University of Groningen's CO-FUND Rosalind Franklin program. PD also warmly thanks the Institute for Advanced Study (IAS) Princeton, where a part of this work was carried out, for their generous hospitality and support through the Bershadsky Fund.

\end{acknowledgements}

%
\bibliographystyle{aa} 
\bibliography{fgasz7.bib} 

\begin{thebibliography}{132}
\expandafter\ifx\csname natexlab\endcsname\relax\def\natexlab#1{#1}\fi

\bibitem[{{Algera} {et~al.}(2023{\natexlab{a}}){Algera}, {Inami}, {Sommovigo},
  {Fudamoto}, {Schneider}, {Graziani}, {Dayal}, {Bouwens}, {Aravena}, {da
  Cunha}, {Ferrara}, {Hygate}, {van Leeuwen}, {De Looze}, {Palla},
  {Pallottini}, {Smit}, {Stefanon}, {Topping}, \& {van der Werf}}]{algera23b}
{Algera}, H., {Inami}, H., {Sommovigo}, L., {et~al.} 2023{\natexlab{a}}, arXiv
  e-prints, arXiv:2301.09659

\bibitem[{{Algera} {et~al.}(2023{\natexlab{b}}){Algera}, {Inami}, {Oesch},
  {Sommovigo}, {Bouwens}, {Topping}, {Schouws}, {Stefanon}, {Stark}, {Aravena},
  {Barrufet}, {da Cunha}, {Dayal}, {Endsley}, {Ferrara}, {Fudamoto},
  {Gonzalez}, {Graziani}, {Hodge}, {Hygate}, {de Looze}, {Nanayakkara},
  {Schneider}, \& {van der Werf}}]{algera23a}
{Algera}, H. S.~B., {Inami}, H., {Oesch}, P.~A., {et~al.} 2023{\natexlab{b}},
  \mnras, 518, 6142

\bibitem[{{Aravena} {et~al.}(2020){Aravena}, {Boogaard},
  {G{\'o}nzalez-L{\'o}pez}, {Decarli}, {Walter}, {Carilli}, {Smail}, {Weiss},
  {Assef}, {Bauer}, {Bouwens}, {Cortes}, {Cox}, {da Cunha}, {Daddi},
  {D{\'\i}az-Santos}, {Inami}, {Ivison}, {Novak}, {Popping}, {Riechers}, {van
  der Werf}, \& {Wagg}}]{aravena20}
{Aravena}, M., {Boogaard}, L., {G{\'o}nzalez-L{\'o}pez}, J., {et~al.} 2020,
  \apj, 901, 79

\bibitem[{{Aravena} {et~al.}(2012){Aravena}, {Carilli}, {Salvato}, {Tanaka},
  {Lentati}, {Schinnerer}, {Walter}, {Riechers}, {Smolci{\'c}}, {Capak},
  {Aussel}, {Bertoldi}, {Chapman}, {Farrah}, {Finoguenov}, {Le Floc'h}, {Lutz},
  {Magdis}, {Oliver}, {Riguccini}, {Berta}, {Magnelli}, \& {Pozzi}}]{aravena12}
{Aravena}, M., {Carilli}, C.~L., {Salvato}, M., {et~al.} 2012, \mnras, 426, 258

\bibitem[{{Aravena} {et~al.}(2019){Aravena}, {Decarli},
  {G{\'o}nzalez-L{\'o}pez}, {Boogaard}, {Walter}, {Carilli}, {Popping},
  {Weiss}, {Assef}, {Bacon}, {Bauer}, {Bertoldi}, {Bouwens}, {Contini},
  {Cortes}, {Cox}, {da Cunha}, {Daddi}, {D{\'\i}az-Santos}, {Elbaz}, {Hodge},
  {Inami}, {Ivison}, {Le F{\`e}vre}, {Magnelli}, {Oesch}, {Riechers}, {Smail},
  {Somerville}, {Swinbank}, {Uzgil}, {van der Werf}, {Wagg}, \&
  {Wisotzki}}]{aravena19}
{Aravena}, M., {Decarli}, R., {G{\'o}nzalez-L{\'o}pez}, J., {et~al.} 2019,
  \apj, 882, 136

\bibitem[{{Aravena} {et~al.}(2016{\natexlab{a}}){Aravena}, {Decarli}, {Walter},
  {Bouwens}, {Oesch}, {Carilli}, {Bauer}, {Da Cunha}, {Daddi},
  {G{\'o}nzalez-L{\'o}pez}, {Ivison}, {Riechers}, {Smail}, {Swinbank}, {Weiss},
  {Anguita}, {Bacon}, {Bell}, {Bertoldi}, {Cortes}, {Cox}, {Hodge}, {Ibar},
  {Inami}, {Infante}, {Karim}, {Magnelli}, {Ota}, {Popping}, {van der Werf},
  {Wagg}, \& {Fudamoto}}]{aravena16c}
{Aravena}, M., {Decarli}, R., {Walter}, F., {et~al.} 2016{\natexlab{a}}, \apj,
  833, 71

\bibitem[{{Aravena} {et~al.}(2016{\natexlab{b}}){Aravena}, {Decarli}, {Walter},
  {Da Cunha}, {Bauer}, {Carilli}, {Daddi}, {Elbaz}, {Ivison}, {Riechers},
  {Smail}, {Swinbank}, {Weiss}, {Anguita}, {Assef}, {Bell}, {Bertoldi},
  {Bacon}, {Bouwens}, {Cortes}, {Cox}, {G{\'o}nzalez-L{\'o}pez}, {Hodge},
  {Ibar}, {Inami}, {Infante}, {Karim}, {Le Le F{\`e}vre}, {Magnelli}, {Ota},
  {Popping}, {Sheth}, {van der Werf}, \& {Wagg}}]{aravena16b}
{Aravena}, M., {Decarli}, R., {Walter}, F., {et~al.} 2016{\natexlab{b}}, \apj,
  833, 68

\bibitem[{{Aravena} {et~al.}(2016{\natexlab{c}}){Aravena}, {Spilker},
  {Bethermin}, {Bothwell}, {Chapman}, {de Breuck}, {Furstenau},
  {G{\'o}nzalez-L{\'o}pez}, {Greve}, {Litke}, {Ma}, {Malkan}, {Marrone},
  {Murphy}, {Stark}, {Strandet}, {Vieira}, {Weiss}, {Welikala}, {Wong}, \&
  {Collier}}]{aravena16a}
{Aravena}, M., {Spilker}, J.~S., {Bethermin}, M., {et~al.} 2016{\natexlab{c}},
  \mnras, 457, 4406

\bibitem[{{Barrufet} {et~al.}(2023){Barrufet}, {Oesch}, {Bouwens}, {Inami},
  {Sommovigo}, {Algera}, {da Cunha}, {Aravena}, {Dayal}, {Ferrara}, {Fudamoto},
  {Gonzalez}, {Graziani}, {Hygate}, {de Looze}, {Nanayakkara}, {Pallottini},
  {Schneider}, {Stefanon}, {Topping}, \& {van der Werf}}]{barrufet23}
{Barrufet}, L., {Oesch}, P.~A., {Bouwens}, R., {et~al.} 2023, \mnras, 522, 3926

\bibitem[{{Belli} {et~al.}(2021){Belli}, {Contursi}, {Genzel}, {Tacconi},
  {F{\"o}rster-Schreiber}, {Lutz}, {Combes}, {Neri}, {Garc{\'\i}a-Burillo},
  {Schuster}, {Herrera-Camus}, {Tadaki}, {Davies}, {Davies}, {Johnson}, {Lee},
  {Leja}, {Nelson}, {Price}, {Shangguan}, {Shimizu}, {Tacchella}, \&
  {{\"U}bler}}]{belli21}
{Belli}, S., {Contursi}, A., {Genzel}, R., {et~al.} 2021, \apjl, 909, L11

\bibitem[{{B{\'e}thermin} {et~al.}(2020){B{\'e}thermin}, {Fudamoto}, {Ginolfi},
  {Loiacono}, {Khusanova}, {Capak}, {Cassata}, {Faisst}, {Le F{\`e}vre},
  {Schaerer}, {Silverman}, {Yan}, {Amorin}, {Bardelli}, {Boquien}, {Cimatti},
  {Davidzon}, {Dessauges-Zavadsky}, {Fujimoto}, {Gruppioni}, {Hathi}, {Ibar},
  {Jones}, {Koekemoer}, {Lagache}, {Lemaux}, {Moreau}, {Oesch}, {Pozzi},
  {Riechers}, {Talia}, {Toft}, {Vallini}, {Vergani}, {Zamorani}, \&
  {Zucca}}]{bethermin20}
{B{\'e}thermin}, M., {Fudamoto}, Y., {Ginolfi}, M., {et~al.} 2020, \aap, 643,
  A2

\bibitem[{{Bezanson} {et~al.}(2019){Bezanson}, {Spilker}, {Williams},
  {Whitaker}, {Narayanan}, {Weiner}, \& {Franx}}]{bezanson19}
{Bezanson}, R., {Spilker}, J., {Williams}, C.~C., {et~al.} 2019, \apjl, 873,
  L19

\bibitem[{{Bezanson} {et~al.}(2022){Bezanson}, {Spilker}, {Suess}, {Setton},
  {Feldmann}, {Greene}, {Kriek}, {Narayanan}, \& {Verrico}}]{bezanson22}
{Bezanson}, R., {Spilker}, J.~S., {Suess}, K.~A., {et~al.} 2022, \apj, 925, 153

\bibitem[{{Bolatto} {et~al.}(2013){Bolatto}, {Wolfire}, \& {Leroy}}]{bolatto13}
{Bolatto}, A.~D., {Wolfire}, M., \& {Leroy}, A.~K. 2013, \araa, 51, 207

\bibitem[{{Boogaard} {et~al.}(2020){Boogaard}, {van der Werf}, {Weiss},
  {Popping}, {Decarli}, {Walter}, {Aravena}, {Bouwens}, {Riechers},
  {Gonz{\'a}lez-L{\'o}pez}, {Smail}, {Carilli}, {Kaasinen}, {Daddi}, {Cox},
  {D{\'\i}az-Santos}, {Inami}, {Cortes}, \& {Wagg}}]{boogaard20}
{Boogaard}, L.~A., {van der Werf}, P., {Weiss}, A., {et~al.} 2020, \apj, 902,
  109

\bibitem[{{Bothwell} {et~al.}(2013){Bothwell}, {Smail}, {Chapman}, {Genzel},
  {Ivison}, {Tacconi}, {Alaghband-Zadeh}, {Bertoldi}, {Blain}, {Casey}, {Cox},
  {Greve}, {Lutz}, {Neri}, {Omont}, \& {Swinbank}}]{bothwell13}
{Bothwell}, M.~S., {Smail}, I., {Chapman}, S.~C., {et~al.} 2013, \mnras, 429,
  3047

\bibitem[{{Bourne} {et~al.}(2019){Bourne}, {Dunlop}, {Simpson}, {Rowlands},
  {Geach}, \& {McLeod}}]{bourne19}
{Bourne}, N., {Dunlop}, J.~S., {Simpson}, J.~M., {et~al.} 2019, \mnras, 482,
  3135

\bibitem[{{Bouwens} {et~al.}(2021){Bouwens}, {Oesch}, {Stefanon},
  {Illingworth}, {Labb{\'e}}, {Reddy}, {Atek}, {Montes}, {Naidu},
  {Nanayakkara}, {Nelson}, \& {Wilkins}}]{bouwens21}
{Bouwens}, R.~J., {Oesch}, P.~A., {Stefanon}, M., {et~al.} 2021, \aj, 162, 47

\bibitem[{{Bouwens} {et~al.}(2022){Bouwens}, {Smit}, {Schouws}, {Stefanon},
  {Bowler}, {Endsley}, {Gonzalez}, {Inami}, {Stark}, {Oesch}, {Hodge},
  {Aravena}, {da Cunha}, {Dayal}, {Looze}, {Ferrara}, {Fudamoto}, {Graziani},
  {Li}, {Nanayakkara}, {Pallottini}, {Schneider}, {Sommovigo}, {Topping}, {van
  der Werf}, {Algera}, {Barrufet}, {Hygate}, {Labb{\'e}}, {Riechers}, \&
  {Witstok}}]{bouwens22}
{Bouwens}, R.~J., {Smit}, R., {Schouws}, S., {et~al.} 2022, \apj, 931, 160

\bibitem[{{Brinchmann} {et~al.}(2008){Brinchmann}, {Pettini}, \&
  {Charlot}}]{brinchmann08}
{Brinchmann}, J., {Pettini}, M., \& {Charlot}, S. 2008, \mnras, 385, 769

\bibitem[{{Carilli} \& {Blain}(2002)}]{carilli02}
{Carilli}, C.~L. \& {Blain}, A.~W. 2002, \apj, 569, 605

\bibitem[{{Carilli} \& {Walter}(2013)}]{carilli13}
{Carilli}, C.~L. \& {Walter}, F. 2013, \araa, 51, 105

\bibitem[{{Cassata} {et~al.}(2020){Cassata}, {Liu}, {Groves}, {Schinnerer},
  {Ibar}, {Sargent}, {Karim}, {Talia}, {F{\`e}vre}, {Tasca}, {Lemaux},
  {Ribeiro}, {Fiore}, {Romano}, {Mancini}, {Morselli}, {Rodighiero},
  {Rodr{\'\i}guez-Mu{\~n}oz}, {Enia}, \& {Smolcic}}]{cassata20}
{Cassata}, P., {Liu}, D., {Groves}, B., {et~al.} 2020, \apj, 891, 83

\bibitem[{{Chabrier}(2003)}]{chabrier03}
{Chabrier}, G. 2003, \pasp, 115, 763

\bibitem[{{Chevallard} \& {Charlot}(2016)}]{chevallard16}
{Chevallard}, J. \& {Charlot}, S. 2016, \mnras, 462, 1415

\bibitem[{{da Cunha} {et~al.}(2013){da Cunha}, {Groves}, {Walter}, {Decarli},
  {Weiss}, {Bertoldi}, {Carilli}, {Daddi}, {Elbaz}, {Ivison}, {Maiolino},
  {Riechers}, {Rix}, {Sargent}, \& {Smail}}]{dacunha13}
{da Cunha}, E., {Groves}, B., {Walter}, F., {et~al.} 2013, \apj, 766, 13

\bibitem[{{Daddi} {et~al.}(2010){Daddi}, {Bournaud}, {Walter}, {Dannerbauer},
  {Carilli}, {Dickinson}, {Elbaz}, {Morrison}, {Riechers}, {Onodera}, {Salmi},
  {Krips}, \& {Stern}}]{daddi10}
{Daddi}, E., {Bournaud}, F., {Walter}, F., {et~al.} 2010, \apj, 713, 686

\bibitem[{{Daddi} {et~al.}(2008){Daddi}, {Dannerbauer}, {Elbaz}, {Dickinson},
  {Morrison}, {Stern}, \& {Ravindranath}}]{daddi08}
{Daddi}, E., {Dannerbauer}, H., {Elbaz}, D., {et~al.} 2008, \apjl, 673, L21

\bibitem[{{Daddi} {et~al.}(2007){Daddi}, {Dickinson}, {Morrison}, {Chary},
  {Cimatti}, {Elbaz}, {Frayer}, {Renzini}, {Pope}, {Alexander}, {Bauer},
  {Giavalisco}, {Huynh}, {Kurk}, \& {Mignoli}}]{daddi07}
{Daddi}, E., {Dickinson}, M., {Morrison}, G., {et~al.} 2007, \apj, 670, 156

\bibitem[{{Dayal} {et~al.}(2022){Dayal}, {Ferrara}, {Sommovigo}, {Bouwens},
  {Oesch}, {Smit}, {Gonzalez}, {Schouws}, {Stefanon}, {Kobayashi}, {Bremer},
  {Algera}, {Aravena}, {Bowler}, {da Cunha}, {Fudamoto}, {Graziani}, {Hodge},
  {Inami}, {De Looze}, {Pallottini}, {Riechers}, {Schneider}, {Stark}, \&
  {Endsley}}]{dayal22}
{Dayal}, P., {Ferrara}, A., {Sommovigo}, L., {et~al.} 2022, \mnras, 512, 989

\bibitem[{{De Looze} {et~al.}(2014){De Looze}, {Cormier}, {Lebouteiller},
  {Madden}, {Baes}, {Bendo}, {Boquien}, {Boselli}, {Clements}, {Cortese},
  {Cooray}, {Galametz}, {Galliano}, {Graci{\'a}-Carpio}, {Isaak}, {Karczewski},
  {Parkin}, {Pellegrini}, {R{\'e}my-Ruyer}, {Spinoglio}, {Smith}, \&
  {Sturm}}]{delooze14}
{De Looze}, I., {Cormier}, D., {Lebouteiller}, V., {et~al.} 2014, \aap, 568,
  A62

\bibitem[{{Decarli} {et~al.}(2020){Decarli}, {Aravena}, {Boogaard}, {Carilli},
  {Gonz{\'a}lez-L{\'o}pez}, {Walter}, {Cortes}, {Cox}, {da Cunha}, {Daddi},
  {D{\'\i}az-Santos}, {Hodge}, {Inami}, {Neeleman}, {Novak}, {Oesch},
  {Popping}, {Riechers}, {Smail}, {Uzgil}, {van der Werf}, {Wagg}, \&
  {Weiss}}]{decarli20}
{Decarli}, R., {Aravena}, M., {Boogaard}, L., {et~al.} 2020, \apj, 902, 110

\bibitem[{{Decarli} {et~al.}(2016){Decarli}, {Walter}, {Aravena}, {Carilli},
  {Bouwens}, {da Cunha}, {Daddi}, {Ivison}, {Popping}, {Riechers}, {Smail},
  {Swinbank}, {Weiss}, {Anguita}, {Assef}, {Bauer}, {Bell}, {Bertoldi},
  {Chapman}, {Colina}, {Cortes}, {Cox}, {Dickinson}, {Elbaz},
  {G{\'o}nzalez-L{\'o}pez}, {Ibar}, {Infante}, {Hodge}, {Karim}, {Le Fevre},
  {Magnelli}, {Neri}, {Oesch}, {Ota}, {Rix}, {Sargent}, {Sheth}, {van der Wel},
  {van der Werf}, \& {Wagg}}]{decarli16a}
{Decarli}, R., {Walter}, F., {Aravena}, M., {et~al.} 2016, \apj, 833, 69

\bibitem[{{Decarli} {et~al.}(2014){Decarli}, {Walter}, {Carilli}, {Riechers},
  {Cox}, {Neri}, {Aravena}, {Bell}, {Bertoldi}, {Colombo}, {Da Cunha}, {Daddi},
  {Dickinson}, {Downes}, {Ellis}, {Lentati}, {Maiolino}, {Menten}, {Rix},
  {Sargent}, {Stark}, {Weiner}, \& {Weiss}}]{decarli14}
{Decarli}, R., {Walter}, F., {Carilli}, C., {et~al.} 2014, \apj, 782, 78

\bibitem[{{Decarli} {et~al.}(2019){Decarli}, {Walter},
  {G{\'o}nzalez-L{\'o}pez}, {Aravena}, {Boogaard}, {Carilli}, {Cox}, {Daddi},
  {Popping}, {Riechers}, {Uzgil}, {Weiss}, {Assef}, {Bacon}, {Bauer},
  {Bertoldi}, {Bouwens}, {Contini}, {Cortes}, {da Cunha}, {D{\'\i}az-Santos},
  {Elbaz}, {Inami}, {Hodge}, {Ivison}, {Le F{\`e}vre}, {Magnelli}, {Novak},
  {Oesch}, {Rix}, {Sargent}, {Smail}, {Swinbank}, {Somerville}, {van der Werf},
  {Wagg}, \& {Wisotzki}}]{decarli19}
{Decarli}, R., {Walter}, F., {G{\'o}nzalez-L{\'o}pez}, J., {et~al.} 2019, \apj,
  882, 138

\bibitem[{{Dessauges-Zavadsky} {et~al.}(2020){Dessauges-Zavadsky}, {Ginolfi},
  {Pozzi}, {B{\'e}thermin}, {Le F{\`e}vre}, {Fujimoto}, {Silverman}, {Jones},
  {Vallini}, {Schaerer}, {Faisst}, {Khusanova}, {Fudamoto}, {Cassata},
  {Loiacono}, {Capak}, {Yan}, {Amorin}, {Bardelli}, {Boquien}, {Cimatti},
  {Gruppioni}, {Hathi}, {Ibar}, {Koekemoer}, {Lemaux}, {Narayanan}, {Oesch},
  {Rodighiero}, {Romano}, {Talia}, {Toft}, {Vergani}, {Zamorani}, \&
  {Zucca}}]{dessauges20}
{Dessauges-Zavadsky}, M., {Ginolfi}, M., {Pozzi}, F., {et~al.} 2020, \aap, 643,
  A5

\bibitem[{{Dessauges-Zavadsky} {et~al.}(2017){Dessauges-Zavadsky}, {Zamojski},
  {Rujopakarn}, {Richard}, {Sklias}, {Schaerer}, {Combes}, {Ebeling}, {Rawle},
  {Egami}, {Boone}, {Cl{\'e}ment}, {Kneib}, {Nyland}, \& {Walth}}]{dessauges17}
{Dessauges-Zavadsky}, M., {Zamojski}, M., {Rujopakarn}, W., {et~al.} 2017,
  \aap, 605, A81

\bibitem[{{Dessauges-Zavadsky} {et~al.}(2015){Dessauges-Zavadsky}, {Zamojski},
  {Schaerer}, {Combes}, {Egami}, {Swinbank}, {Richard}, {Sklias}, {Rawle},
  {Rex}, {Kneib}, {Boone}, \& {Blain}}]{dessauges15}
{Dessauges-Zavadsky}, M., {Zamojski}, M., {Schaerer}, D., {et~al.} 2015, \aap,
  577, A50

\bibitem[{{Di Cesare} {et~al.}(2023){Di Cesare}, {Graziani}, {Schneider},
  {Ginolfi}, {Venditti}, {Santini}, \& {Hunt}}]{dicesare23}
{Di Cesare}, C., {Graziani}, L., {Schneider}, R., {et~al.} 2023, \mnras, 519,
  4632

\bibitem[{{Downes} \& {Solomon}(1998)}]{downes98}
{Downes}, D. \& {Solomon}, P.~M. 1998, \apj, 507, 615

\bibitem[{{Dunlop} {et~al.}(2017){Dunlop}, {McLure}, {Biggs}, {Geach},
  {Micha{\l}owski}, {Ivison}, {Rujopakarn}, {van Kampen}, {Kirkpatrick},
  {Pope}, {Scott}, {Swinbank}, {Targett}, {Aretxaga}, {Austermann}, {Best},
  {Bruce}, {Chapin}, {Charlot}, {Cirasuolo}, {Coppin}, {Ellis}, {Finkelstein},
  {Hayward}, {Hughes}, {Ibar}, {Jagannathan}, {Khochfar}, {Koprowski},
  {Narayanan}, {Nyland}, {Papovich}, {Peacock}, {Rieke}, {Robertson},
  {Vernstrom}, {Werf}, {Wilson}, \& {Yun}}]{dunlop17}
{Dunlop}, J.~S., {McLure}, R.~J., {Biggs}, A.~D., {et~al.} 2017, \mnras, 466,
  861

\bibitem[{{Elbaz} {et~al.}(2007){Elbaz}, {Daddi}, {Le Borgne}, {Dickinson},
  {Alexander}, {Chary}, {Starck}, {Brandt}, {Kitzbichler}, {MacDonald},
  {Nonino}, {Popesso}, {Stern}, \& {Vanzella}}]{elbaz07}
{Elbaz}, D., {Daddi}, E., {Le Borgne}, D., {et~al.} 2007, \aap, 468, 33

\bibitem[{{Elbaz} {et~al.}(2011){Elbaz}, {Dickinson}, {Hwang},
  {D{\'\i}az-Santos}, {Magdis}, {Magnelli}, {Le Borgne}, {Galliano},
  {Pannella}, {Chanial}, {Armus}, {Charmandaris}, {Daddi}, {Aussel}, {Popesso},
  {Kartaltepe}, {Altieri}, {Valtchanov}, {Coia}, {Dannerbauer}, {Dasyra},
  {Leiton}, {Mazzarella}, {Alexander}, {Buat}, {Burgarella}, {Chary}, {Gilli},
  {Ivison}, {Juneau}, {Le Floc'h}, {Lutz}, {Morrison}, {Mullaney}, {Murphy},
  {Pope}, {Scott}, {Brodwin}, {Calzetti}, {Cesarsky}, {Charlot}, {Dole},
  {Eisenhardt}, {Ferguson}, {F{\"o}rster Schreiber}, {Frayer}, {Giavalisco},
  {Huynh}, {Koekemoer}, {Papovich}, {Reddy}, {Surace}, {Teplitz}, {Yun}, \&
  {Wilson}}]{elbaz11}
{Elbaz}, D., {Dickinson}, M., {Hwang}, H.~S., {et~al.} 2011, \aap, 533, A119

\bibitem[{{Endsley} {et~al.}(2022){Endsley}, {Stark}, {Bouwens}, {Schouws},
  {Smit}, {Stefanon}, {Inami}, {Bowler}, {Oesch}, {Gonzalez}, {Aravena}, {da
  Cunha}, {Dayal}, {Ferrara}, {Graziani}, {Nanayakkara}, {Pallottini},
  {Schneider}, {Sommovigo}, {Topping}, {van der Werf}, \& {Hutter}}]{endsley22}
{Endsley}, R., {Stark}, D.~P., {Bouwens}, R.~J., {et~al.} 2022, \mnras, 517,
  5642

\bibitem[{{Ferrara} {et~al.}(2022){Ferrara}, {Sommovigo}, {Dayal},
  {Pallottini}, {Bouwens}, {Gonzalez}, {Inami}, {Smit}, {Bowler}, {Endsley},
  {Oesch}, {Schouws}, {Stark}, {Stefanon}, {Aravena}, {da Cunha}, {De Looze},
  {Fudamoto}, {Graziani}, {Hodge}, {Riechers}, {Schneider}, {Algera},
  {Barrufet}, {Hygate}, {Labb{\'e}}, {Li}, {Nanayakkara}, {Topping}, \& {van
  der Werf}}]{ferrara22}
{Ferrara}, A., {Sommovigo}, L., {Dayal}, P., {et~al.} 2022, \mnras, 512, 58

\bibitem[{{Ferrara} {et~al.}(2019){Ferrara}, {Vallini}, {Pallottini},
  {Gallerani}, {Carniani}, {Kohandel}, {Decataldo}, \& {Behrens}}]{ferrara19}
{Ferrara}, A., {Vallini}, L., {Pallottini}, A., {et~al.} 2019, \mnras, 489, 1

\bibitem[{{Franco} {et~al.}(2020){Franco}, {Elbaz}, {Zhou}, {Magnelli},
  {Schreiber}, {Ciesla}, {Dickinson}, {Nagar}, {Magdis}, {Alexander},
  {B{\'e}thermin}, {Demarco}, {Daddi}, {Wang}, {Mullaney}, {Sargent}, {Inami},
  {Shu}, {Bournaud}, {Chary}, {Coogan}, {Ferguson}, {Finkelstein},
  {Giavalisco}, {G{\'o}mez-Guijarro}, {Iono}, {Juneau}, {Lagache}, {Lin},
  {Motohara}, {Okumura}, {Pannella}, {Papovich}, {Pope}, {Rujopakarn},
  {Silverman}, \& {Xiao}}]{franco20}
{Franco}, M., {Elbaz}, D., {Zhou}, L., {et~al.} 2020, \aap, 643, A30

\bibitem[{{Freundlich} {et~al.}(2019){Freundlich}, {Combes}, {Tacconi},
  {Genzel}, {Garcia-Burillo}, {Neri}, {Contini}, {Bolatto}, {Lilly},
  {Salom{\'e}}, {Bicalho}, {Boissier}, {Boone}, {Bouch{\'e}}, {Bournaud},
  {Burkert}, {Carollo}, {Cooper}, {Cox}, {Feruglio}, {F{\"o}rster Schreiber},
  {Juneau}, {Lippa}, {Lutz}, {Naab}, {Renzini}, {Saintonge}, {Sternberg},
  {Walter}, {Weiner}, {Wei{\ss}}, \& {Wuyts}}]{freundlich19}
{Freundlich}, J., {Combes}, F., {Tacconi}, L.~J., {et~al.} 2019, \aap, 622,
  A105

\bibitem[{{Fudamoto} {et~al.}(2017){Fudamoto}, {Ivison}, {Oteo}, {Krips},
  {Zhang}, {Weiss}, {Dannerbauer}, {Omont}, {Chapman}, {Christensen},
  {Arumugam}, {Bertoldi}, {Bremer}, {Clements}, {Dunne}, {Eales}, {Greenslade},
  {Maddox}, {Martinez-Navajas}, {Michalowski}, {P{\'e}rez-Fournon}, {Riechers},
  {Simpson}, {Stalder}, {Valiante}, \& {van der Werf}}]{fudamoto17}
{Fudamoto}, Y., {Ivison}, R.~J., {Oteo}, I., {et~al.} 2017, \mnras, 472, 2028

\bibitem[{{Fudamoto} {et~al.}(2021){Fudamoto}, {Oesch}, {Schouws}, {Stefanon},
  {Smit}, {Bouwens}, {Bowler}, {Endsley}, {Gonzalez}, {Inami}, {Labbe},
  {Stark}, {Aravena}, {Barrufet}, {da Cunha}, {Dayal}, {Ferrara}, {Graziani},
  {Hodge}, {Hutter}, {Li}, {De Looze}, {Nanayakkara}, {Pallottini}, {Riechers},
  {Schneider}, {Ucci}, {van der Werf}, \& {White}}]{fudamoto21}
{Fudamoto}, Y., {Oesch}, P.~A., {Schouws}, S., {et~al.} 2021, \nat, 597, 489

\bibitem[{{Fudamoto} {et~al.}(2022){Fudamoto}, {Smit}, {Bowler}, {Oesch},
  {Bouwens}, {Stefanon}, {Inami}, {Endsley}, {Gonzalez}, {Schouws}, {Stark},
  {Algera}, {Aravena}, {Barrufet}, {da Cunha}, {Dayal}, {Ferrara}, {Graziani},
  {Hodge}, {Hygate}, {Inoue}, {Nanayakkara}, {Pallottini}, {Pizzati},
  {Schneider}, {Sommovigo}, {Sugahara}, {Topping}, {van der Werf}, {Bethermin},
  {Cassata}, {Dessauges-Zavadsky}, {Ibar}, {Faisst}, {Fujimoto}, {Ginolfi},
  {Hathi}, {Jones}, {Pozzi}, \& {Schaerer}}]{fudamoto22}
{Fudamoto}, Y., {Smit}, R., {Bowler}, R.~A.~A., {et~al.} 2022, \apj, 934, 144

\bibitem[{{Fujimoto} {et~al.}(2020){Fujimoto}, {Silverman}, {Bethermin},
  {Ginolfi}, {Jones}, {Le F{\`e}vre}, {Dessauges-Zavadsky}, {Rujopakarn},
  {Faisst}, {Fudamoto}, {Cassata}, {Morselli}, {Maiolino}, {Schaerer}, {Capak},
  {Yan}, {Vallini}, {Toft}, {Loiacono}, {Zamorani}, {Talia}, {Narayanan},
  {Hathi}, {Lemaux}, {Boquien}, {Amorin}, {Ibar}, {Koekemoer},
  {M{\'e}ndez-Hern{\'a}ndez}, {Bardelli}, {Vergani}, {Zucca}, {Romano}, \&
  {Cimatti}}]{fujimoto20}
{Fujimoto}, S., {Silverman}, J.~D., {Bethermin}, M., {et~al.} 2020, \apj, 900,
  1

\bibitem[{{Genzel} {et~al.}(2015){Genzel}, {Tacconi}, {Lutz}, {Saintonge},
  {Berta}, {Magnelli}, {Combes}, {Garc{\'\i}a-Burillo}, {Neri}, {Bolatto},
  {Contini}, {Lilly}, {Boissier}, {Boone}, {Bouch{\'e}}, {Bournaud}, {Burkert},
  {Carollo}, {Colina}, {Cooper}, {Cox}, {Feruglio}, {F{\"o}rster Schreiber},
  {Freundlich}, {Gracia-Carpio}, {Juneau}, {Kovac}, {Lippa}, {Naab}, {Salome},
  {Renzini}, {Sternberg}, {Walter}, {Weiner}, {Weiss}, \& {Wuyts}}]{genzel15}
{Genzel}, R., {Tacconi}, L.~J., {Lutz}, D., {et~al.} 2015, \apj, 800, 20

\bibitem[{{Gonz{\'a}lez-L{\'o}pez} {et~al.}(2019){Gonz{\'a}lez-L{\'o}pez},
  {Decarli}, {Pavesi}, {Walter}, {Aravena}, {Carilli}, {Boogaard}, {Popping},
  {Weiss}, {Assef}, {Bauer}, {Bertoldi}, {Bouwens}, {Contini}, {Cortes}, {Cox},
  {da Cunha}, {Daddi}, {D{\'\i}az-Santos}, {Inami}, {Hodge}, {Ivison}, {Le
  F{\`e}vre}, {Magnelli}, {Oesch}, {Riechers}, {Rix}, {Smail}, {Swinbank},
  {Somerville}, {Uzgil}, \& {van der Werf}}]{gonzalezlopez19}
{Gonz{\'a}lez-L{\'o}pez}, J., {Decarli}, R., {Pavesi}, R., {et~al.} 2019, \apj,
  882, 139

\bibitem[{{Gonz{\'a}lez-L{\'o}pez} {et~al.}(2020){Gonz{\'a}lez-L{\'o}pez},
  {Novak}, {Decarli}, {Walter}, {Aravena}, {Carilli}, {Boogaard}, {Popping},
  {Weiss}, {Assef}, {Bauer}, {Bouwens}, {Cortes}, {Cox}, {Daddi}, {Cunha},
  {D{\'\i}az-Santos}, {Ivison}, {Magnelli}, {Riechers}, {Smail}, {van der
  Werf}, \& {Wagg}}]{gonzalezlopez20}
{Gonz{\'a}lez-L{\'o}pez}, J., {Novak}, M., {Decarli}, R., {et~al.} 2020, \apj,
  897, 91

\bibitem[{{Gowardhan} {et~al.}(2019){Gowardhan}, {Riechers}, {Pavesi}, {Daddi},
  {Dannerbauer}, \& {Neri}}]{gowardhan19}
{Gowardhan}, A., {Riechers}, D., {Pavesi}, R., {et~al.} 2019, \apj, 875, 6

\bibitem[{{Graziani} {et~al.}(2020){Graziani}, {Schneider}, {Ginolfi}, {Hunt},
  {Maio}, {Glatzle}, \& {Ciardi}}]{graziani20}
{Graziani}, L., {Schneider}, R., {Ginolfi}, M., {et~al.} 2020, \mnras, 494,
  1071

\bibitem[{{Grogin} {et~al.}(2011){Grogin}, {Kocevski}, {Faber}, {Ferguson},
  {Koekemoer}, {Riess}, {Acquaviva}, {Alexander}, {Almaini}, {Ashby}, {Barden},
  {Bell}, {Bournaud}, {Brown}, {Caputi}, {Casertano}, {Cassata}, {Castellano},
  {Challis}, {Chary}, {Cheung}, {Cirasuolo}, {Conselice}, {Roshan Cooray},
  {Croton}, {Daddi}, {Dahlen}, {Dav{\'e}}, {de Mello}, {Dekel}, {Dickinson},
  {Dolch}, {Donley}, {Dunlop}, {Dutton}, {Elbaz}, {Fazio}, {Filippenko},
  {Finkelstein}, {Fontana}, {Gardner}, {Garnavich}, {Gawiser}, {Giavalisco},
  {Grazian}, {Guo}, {Hathi}, {H{\"a}ussler}, {Hopkins}, {Huang}, {Huang},
  {Jha}, {Kartaltepe}, {Kirshner}, {Koo}, {Lai}, {Lee}, {Li}, {Lotz}, {Lucas},
  {Madau}, {McCarthy}, {McGrath}, {McIntosh}, {McLure}, {Mobasher},
  {Moustakas}, {Mozena}, {Nandra}, {Newman}, {Niemi}, {Noeske}, {Papovich},
  {Pentericci}, {Pope}, {Primack}, {Rajan}, {Ravindranath}, {Reddy}, {Renzini},
  {Rix}, {Robaina}, {Rodney}, {Rosario}, {Rosati}, {Salimbeni}, {Scarlata},
  {Siana}, {Simard}, {Smidt}, {Somerville}, {Spinrad}, {Straughn}, {Strolger},
  {Telford}, {Teplitz}, {Trump}, {van der Wel}, {Villforth}, {Wechsler},
  {Weiner}, {Wiklind}, {Wild}, {Wilson}, {Wuyts}, {Yan}, \& {Yun}}]{grogin11}
{Grogin}, N.~A., {Kocevski}, D.~D., {Faber}, S.~M., {et~al.} 2011, \apjs, 197,
  35

\bibitem[{{Hatsukade} {et~al.}(2018){Hatsukade}, {Kohno}, {Yamaguchi},
  {Umehata}, {Ao}, {Aretxaga}, {Caputi}, {Dunlop}, {Egami}, {Espada},
  {Fujimoto}, {Hayatsu}, {Hughes}, {Ikarashi}, {Iono}, {Ivison}, {Kawabe},
  {Kodama}, {Lee}, {Matsuda}, {Nakanishi}, {Ohta}, {Ouchi}, {Rujopakarn},
  {Suzuki}, {Tamura}, {Ueda}, {Wang}, {Wang}, {Wilson}, {Yoshimura}, \&
  {Yun}}]{hatsukade18}
{Hatsukade}, B., {Kohno}, K., {Yamaguchi}, Y., {et~al.} 2018, \pasj, 70, 105

\bibitem[{{Heintz} {et~al.}(2022){Heintz}, {Oesch}, {Aravena}, {Bouwens},
  {Dayal}, {Ferrara}, {Fudamoto}, {Graziani}, {Inami}, {Sommovigo}, {Smit},
  {Stefanon}, {Topping}, {Pallottini}, \& {van der Werf}}]{heintz22}
{Heintz}, K.~E., {Oesch}, P.~A., {Aravena}, M., {et~al.} 2022, \apjl, 934, L27

\bibitem[{{Heintz} \& {Watson}(2020)}]{heintz20}
{Heintz}, K.~E. \& {Watson}, D. 2020, \apjl, 889, L7

\bibitem[{{Heintz} {et~al.}(2021){Heintz}, {Watson}, {Oesch}, {Narayanan}, \&
  {Madden}}]{heintz21}
{Heintz}, K.~E., {Watson}, D., {Oesch}, P.~A., {Narayanan}, D., \& {Madden},
  S.~C. 2021, \apj, 922, 147

\bibitem[{{Herrera-Camus} {et~al.}(2021){Herrera-Camus}, {F{\"o}rster
  Schreiber}, {Genzel}, {Tacconi}, {Bolatto}, {Davies}, {Fisher}, {Lutz},
  {Naab}, {Shimizu}, {Tadaki}, \& {{\"U}bler}}]{herreracamus21}
{Herrera-Camus}, R., {F{\"o}rster Schreiber}, N., {Genzel}, R., {et~al.} 2021,
  \aap, 649, A31

\bibitem[{{Hodge} \& {da Cunha}(2020)}]{hodge20}
{Hodge}, J.~A. \& {da Cunha}, E. 2020, Royal Society Open Science, 7, 200556

\bibitem[{{Hughes} {et~al.}(2017){Hughes}, {Ibar}, {Villanueva}, {Aravena},
  {Baes}, {Bourne}, {Cooray}, {Dunne}, {Dye}, {Eales}, {Furlanetto},
  {Herrera-Camus}, {Ivison}, {van Kampen}, {Lara-L{\'o}pez}, {Maddox},
  {Micha{\l}owski}, {Smith}, {Valiante}, {van der Werf}, \& {Xue}}]{hughes17}
{Hughes}, T.~M., {Ibar}, E., {Villanueva}, V., {et~al.} 2017, \aap, 602, A49

\bibitem[{{Hygate} {et~al.}(2023){Hygate}, {Hodge}, {da Cunha}, {Rybak},
  {Schouws}, {Inami}, {Stefanon}, {Graziani}, {Schneider}, {Dayal}, {Bouwens},
  {Smit}, {Bowler}, {Endsley}, {Gonzalez}, {Oesch}, {Stark}, {Algera},
  {Aravena}, {Barrufet}, {Ferrara}, {Fudamoto}, {Hilhorst}, {De Looze},
  {Nanayakkara}, {Pallottini}, {Riechers}, {Sommovigo}, {Topping}, \& {van der
  Werf}}]{hygate23}
{Hygate}, A.~P.~S., {Hodge}, J.~A., {da Cunha}, E., {et~al.} 2023, \mnras
  [\eprint[arXiv]{2304.09206}]

\bibitem[{{Inami} {et~al.}(2022){Inami}, {Algera}, {Schouws}, {Sommovigo},
  {Bouwens}, {Smit}, {Stefanon}, {Bowler}, {Endsley}, {Ferrara}, {Oesch},
  {Stark}, {Aravena}, {Barrufet}, {da Cunha}, {Dayal}, {De Looze}, {Fudamoto},
  {Gonzalez}, {Graziani}, {Hodge}, {Hygate}, {Nanayakkara}, {Pallottini},
  {Riechers}, {Schneider}, {Topping}, \& {van der Werf}}]{inami22}
{Inami}, H., {Algera}, H. S.~B., {Schouws}, S., {et~al.} 2022, \mnras, 515,
  3126

\bibitem[{{Iyer} {et~al.}(2018){Iyer}, {Gawiser}, {Dav{\'e}}, {Davis},
  {Finkelstein}, {Kodra}, {Koekemoer}, {Kurczynski}, {Newman}, {Pacifici}, \&
  {Somerville}}]{iyer18}
{Iyer}, K., {Gawiser}, E., {Dav{\'e}}, R., {et~al.} 2018, \apj, 866, 120

\bibitem[{{Jarvis} {et~al.}(2013){Jarvis}, {Bonfield}, {Bruce}, {Geach},
  {McAlpine}, {McLure}, {Gonz{\'a}lez-Solares}, {Irwin}, {Lewis}, {Yoldas},
  {Andreon}, {Cross}, {Emerson}, {Dalton}, {Dunlop}, {Hodgkin}, {Le},
  {Karouzos}, {Meisenheimer}, {Oliver}, {Rawlings}, {Simpson}, {Smail},
  {Smith}, {Sullivan}, {Sutherland}, {White}, \& {Zwart}}]{jarvis13}
{Jarvis}, M.~J., {Bonfield}, D.~G., {Bruce}, V.~A., {et~al.} 2013, \mnras, 428,
  1281

\bibitem[{{Johnson} {et~al.}(2021){Johnson}, {Leja}, {Conroy}, \&
  {Speagle}}]{johnson21}
{Johnson}, B.~D., {Leja}, J., {Conroy}, C., \& {Speagle}, J.~S. 2021, \apjs,
  254, 22

\bibitem[{{Kaasinen} {et~al.}(2019){Kaasinen}, {Scoville}, {Walter}, {Da
  Cunha}, {Popping}, {Pavesi}, {Darvish}, {Casey}, {Riechers}, \&
  {Glover}}]{kaasinen19}
{Kaasinen}, M., {Scoville}, N., {Walter}, F., {et~al.} 2019, \apj, 880, 15

\bibitem[{{Kartaltepe} {et~al.}(2012){Kartaltepe}, {Dickinson}, {Alexander},
  {Bell}, {Dahlen}, {Elbaz}, {Faber}, {Lotz}, {McIntosh}, {Wiklind}, {Altieri},
  {Aussel}, {Bethermin}, {Bournaud}, {Charmandaris}, {Conselice}, {Cooray},
  {Dannerbauer}, {Dav{\'e}}, {Dunlop}, {Dekel}, {Ferguson}, {Grogin}, {Hwang},
  {Ivison}, {Kocevski}, {Koekemoer}, {Koo}, {Lai}, {Leiton}, {Lucas}, {Lutz},
  {Magdis}, {Magnelli}, {Morrison}, {Mozena}, {Mullaney}, {Newman}, {Pope},
  {Popesso}, {van der Wel}, {Weiner}, \& {Wuyts}}]{kartaltepe12}
{Kartaltepe}, J.~S., {Dickinson}, M., {Alexander}, D.~M., {et~al.} 2012, \apj,
  757, 23

\bibitem[{{Katz} {et~al.}(2019){Katz}, {Galligan}, {Kimm}, {Rosdahl},
  {Haehnelt}, {Blaizot}, {Devriendt}, {Slyz}, {Laporte}, \& {Ellis}}]{katz19}
{Katz}, H., {Galligan}, T.~P., {Kimm}, T., {et~al.} 2019, \mnras, 487, 5902

\bibitem[{{Katz} {et~al.}(2017){Katz}, {Kimm}, {Sijacki}, \&
  {Haehnelt}}]{katz17}
{Katz}, H., {Kimm}, T., {Sijacki}, D., \& {Haehnelt}, M.~G. 2017, \mnras, 468,
  4831

\bibitem[{{Khusanova} {et~al.}(2021){Khusanova}, {Bethermin}, {Le F{\`e}vre},
  {Capak}, {Faisst}, {Schaerer}, {Silverman}, {Cassata}, {Yan}, {Ginolfi},
  {Fudamoto}, {Loiacono}, {Amorin}, {Bardelli}, {Boquien}, {Cimatti},
  {Dessauges-Zavadsky}, {Gruppioni}, {Hathi}, {Jones}, {Koekemoer}, {Lagache},
  {Maiolino}, {Lemaux}, {Oesch}, {Pozzi}, {Riechers}, {Romano}, {Talia},
  {Toft}, {Vergani}, {Zamorani}, \& {Zucca}}]{khusanova21}
{Khusanova}, Y., {Bethermin}, M., {Le F{\`e}vre}, O., {et~al.} 2021, \aap, 649,
  A152

\bibitem[{{Klitsch} {et~al.}(2019){Klitsch}, {P{\'e}roux}, {Zwaan}, {Smail},
  {Nelson}, {Popping}, {Chen}, {Diemer}, {Ivison}, {Allison}, {Muller},
  {Swinbank}, {Hamanowicz}, {Biggs}, \& {Dutta}}]{klitsch19}
{Klitsch}, A., {P{\'e}roux}, C., {Zwaan}, M.~A., {et~al.} 2019, \mnras, 490,
  1220

\bibitem[{{Lagache} {et~al.}(2018){Lagache}, {Cousin}, \&
  {Chatzikos}}]{lagache18}
{Lagache}, G., {Cousin}, M., \& {Chatzikos}, M. 2018, \aap, 609, A130

\bibitem[{{Lawrence} {et~al.}(2007){Lawrence}, {Warren}, {Almaini}, {Edge},
  {Hambly}, {Jameson}, {Lucas}, {Casali}, {Adamson}, {Dye}, {Emerson},
  {Foucaud}, {Hewett}, {Hirst}, {Hodgkin}, {Irwin}, {Lodieu}, {McMahon},
  {Simpson}, {Smail}, {Mortlock}, \& {Folger}}]{lawrence07}
{Lawrence}, A., {Warren}, S.~J., {Almaini}, O., {et~al.} 2007, \mnras, 379,
  1599

\bibitem[{{Le F{\`e}vre} {et~al.}(2020){Le F{\`e}vre}, {B{\'e}thermin},
  {Faisst}, {Jones}, {Capak}, {Cassata}, {Silverman}, {Schaerer}, {Yan},
  {Amorin}, {Bardelli}, {Boquien}, {Cimatti}, {Dessauges-Zavadsky},
  {Giavalisco}, {Hathi}, {Fudamoto}, {Fujimoto}, {Ginolfi}, {Gruppioni},
  {Hemmati}, {Ibar}, {Koekemoer}, {Khusanova}, {Lagache}, {Lemaux}, {Loiacono},
  {Maiolino}, {Mancini}, {Narayanan}, {Morselli}, {M{\'e}ndez-Hern{\`a}ndez},
  {Oesch}, {Pozzi}, {Romano}, {Riechers}, {Scoville}, {Talia}, {Tasca},
  {Thomas}, {Toft}, {Vallini}, {Vergani}, {Walter}, {Zamorani}, \&
  {Zucca}}]{lefevre20}
{Le F{\`e}vre}, O., {B{\'e}thermin}, M., {Faisst}, A., {et~al.} 2020, \aap,
  643, A1

\bibitem[{{Leja} {et~al.}(2019){Leja}, {Carnall}, {Johnson}, {Conroy}, \&
  {Speagle}}]{leja19}
{Leja}, J., {Carnall}, A.~C., {Johnson}, B.~D., {Conroy}, C., \& {Speagle},
  J.~S. 2019, \apj, 876, 3

\bibitem[{{Lenki{\'c}} {et~al.}(2020){Lenki{\'c}}, {Bolatto}, {F{\"o}rster
  Schreiber}, {Tacconi}, {Neri}, {Combes}, {Walter}, {Garc{\'\i}a-Burillo},
  {Genzel}, {Lutz}, \& {Cooper}}]{lenkic20}
{Lenki{\'c}}, L., {Bolatto}, A.~D., {F{\"o}rster Schreiber}, N.~M., {et~al.}
  2020, \aj, 159, 190

\bibitem[{{Leroy} {et~al.}(2008){Leroy}, {Walter}, {Brinks}, {Bigiel}, {de
  Blok}, {Madore}, \& {Thornley}}]{leroy08}
{Leroy}, A.~K., {Walter}, F., {Brinks}, E., {et~al.} 2008, \aj, 136, 2782

\bibitem[{{Liu} {et~al.}(2019){Liu}, {Schinnerer}, {Groves}, {Magnelli},
  {Lang}, {Leslie}, {Jim{\'e}nez-Andrade}, {Riechers}, {Popping}, {Magdis},
  {Daddi}, {Sargent}, {Gao}, {Fudamoto}, {Oesch}, \& {Bertoldi}}]{liu19}
{Liu}, D., {Schinnerer}, E., {Groves}, B., {et~al.} 2019, \apj, 887, 235

\bibitem[{{Madau} \& {Dickinson}(2014)}]{madau14}
{Madau}, P. \& {Dickinson}, M. 2014, \araa, 52, 415

\bibitem[{{Madden} {et~al.}(2020){Madden}, {Cormier}, {Hony}, {Lebouteiller},
  {Abel}, {Galametz}, {De Looze}, {Chevance}, {Polles}, {Lee}, {Galliano},
  {Lambert-Huyghe}, {Hu}, \& {Ramambason}}]{madden20}
{Madden}, S.~C., {Cormier}, D., {Hony}, S., {et~al.} 2020, \aap, 643, A141

\bibitem[{{Magdis} {et~al.}(2017){Magdis}, {Rigopoulou}, {Daddi}, {Bethermin},
  {Feruglio}, {Sargent}, {Dannerbauer}, {Dickinson}, {Elbaz}, {Gomez Guijarro},
  {Huang}, {Toft}, \& {Valentino}}]{magdis17}
{Magdis}, G.~E., {Rigopoulou}, D., {Daddi}, E., {et~al.} 2017, \aap, 603, A93

\bibitem[{{Magnelli} {et~al.}(2020){Magnelli}, {Boogaard}, {Decarli},
  {G{\'o}nzalez-L{\'o}pez}, {Novak}, {Popping}, {Smail}, {Walter}, {Aravena},
  {Assef}, {Bauer}, {Bertoldi}, {Carilli}, {Cortes}, {Cunha}, {Daddi},
  {D{\'\i}az-Santos}, {Inami}, {Ivison}, {F{\`e}vre}, {Oesch}, {Riechers},
  {Rix}, {Sargent}, {Werf}, {Wagg}, \& {Weiss}}]{magnelli20}
{Magnelli}, B., {Boogaard}, L., {Decarli}, R., {et~al.} 2020, \apj, 892, 66

\bibitem[{{McCracken} {et~al.}(2012){McCracken}, {Milvang-Jensen}, {Dunlop},
  {Franx}, {Fynbo}, {Le F{\`e}vre}, {Holt}, {Caputi}, {Goranova}, {Buitrago},
  {Emerson}, {Freudling}, {Hudelot}, {L{\'o}pez-Sanjuan}, {Magnard}, {Mellier},
  {M{\o}ller}, {Nilsson}, {Sutherland}, {Tasca}, \& {Zabl}}]{mccracken12}
{McCracken}, H.~J., {Milvang-Jensen}, B., {Dunlop}, J., {et~al.} 2012, \aap,
  544, A156

\bibitem[{{Molina} {et~al.}(2019){Molina}, {Ibar}, {Smail}, {Swinbank},
  {Villard}, {Escala}, {Sobral}, \& {Hughes}}]{molina19}
{Molina}, J., {Ibar}, E., {Smail}, I., {et~al.} 2019, \mnras, 487, 4856

\bibitem[{{Noeske} {et~al.}(2007){Noeske}, {Weiner}, {Faber}, {Papovich},
  {Koo}, {Somerville}, {Bundy}, {Conselice}, {Newman}, {Schiminovich}, {Le
  Floc'h}, {Coil}, {Rieke}, {Lotz}, {Primack}, {Barmby}, {Cooper}, {Davis},
  {Ellis}, {Fazio}, {Guhathakurta}, {Huang}, {Kassin}, {Martin}, {Phillips},
  {Rich}, {Small}, {Willmer}, \& {Wilson}}]{noeske07}
{Noeske}, K.~G., {Weiner}, B.~J., {Faber}, S.~M., {et~al.} 2007, \apjl, 660,
  L43

\bibitem[{{Olsen} {et~al.}(2017){Olsen}, {Greve}, {Narayanan}, {Thompson},
  {Dav{\'e}}, {Niebla Rios}, \& {Stawinski}}]{olsen17}
{Olsen}, K., {Greve}, T.~R., {Narayanan}, D., {et~al.} 2017, \apj, 846, 105

\bibitem[{{Pallottini} {et~al.}(2017){Pallottini}, {Ferrara}, {Bovino},
  {Vallini}, {Gallerani}, {Maiolino}, \& {Salvadori}}]{pallotini17}
{Pallottini}, A., {Ferrara}, A., {Bovino}, S., {et~al.} 2017, \mnras, 471, 4128

\bibitem[{{Pallottini} {et~al.}(2019){Pallottini}, {Ferrara}, {Decataldo},
  {Gallerani}, {Vallini}, {Carniani}, {Behrens}, {Kohandel}, \&
  {Salvadori}}]{pallotini19}
{Pallottini}, A., {Ferrara}, A., {Decataldo}, D., {et~al.} 2019, \mnras, 487,
  1689

\bibitem[{{Pallottini} {et~al.}(2022){Pallottini}, {Ferrara}, {Gallerani},
  {Behrens}, {Kohandel}, {Carniani}, {Vallini}, {Salvadori}, {Gelli},
  {Sommovigo}, {D'Odorico}, {Di Mascia}, \& {Pizzati}}]{pallotini22}
{Pallottini}, A., {Ferrara}, A., {Gallerani}, S., {et~al.} 2022, \mnras, 513,
  5621

\bibitem[{{Papovich} {et~al.}(2016){Papovich}, {Labb{\'e}}, {Glazebrook},
  {Quadri}, {Bekiaris}, {Dickinson}, {Finkelstein}, {Fisher}, {Inami},
  {Livermore}, {Spitler}, {Straatman}, \& {Tran}}]{papovich16}
{Papovich}, C., {Labb{\'e}}, I., {Glazebrook}, K., {et~al.} 2016, Nature
  Astronomy, 1, 0003

\bibitem[{{Pavesi} {et~al.}(2018){Pavesi}, {Sharon}, {Riechers}, {Hodge},
  {Decarli}, {Walter}, {Carilli}, {Daddi}, {Smail}, {Dickinson}, {Ivison},
  {Sargent}, {da Cunha}, {Aravena}, {Darling}, {Smol{\v{c}}i{\'c}}, {Scoville},
  {Capak}, \& {Wagg}}]{pavesi18}
{Pavesi}, R., {Sharon}, C.~E., {Riechers}, D.~A., {et~al.} 2018, \apj, 864, 49

\bibitem[{{Peng} {et~al.}(2010){Peng}, {Lilly}, {Kova{\v{c}}}, {Bolzonella},
  {Pozzetti}, {Renzini}, {Zamorani}, {Ilbert}, {Knobel}, {Iovino}, {Maier},
  {Cucciati}, {Tasca}, {Carollo}, {Silverman}, {Kampczyk}, {de Ravel},
  {Sanders}, {Scoville}, {Contini}, {Mainieri}, {Scodeggio}, {Kneib}, {Le
  F{\`e}vre}, {Bardelli}, {Bongiorno}, {Caputi}, {Coppa}, {de la Torre},
  {Franzetti}, {Garilli}, {Lamareille}, {Le Borgne}, {Le Brun}, {Mignoli},
  {Perez Montero}, {Pello}, {Ricciardelli}, {Tanaka}, {Tresse}, {Vergani},
  {Welikala}, {Zucca}, {Oesch}, {Abbas}, {Barnes}, {Bordoloi}, {Bottini},
  {Cappi}, {Cassata}, {Cimatti}, {Fumana}, {Hasinger}, {Koekemoer},
  {Leauthaud}, {Maccagni}, {Marinoni}, {McCracken}, {Memeo}, {Meneux}, {Nair},
  {Porciani}, {Presotto}, \& {Scaramella}}]{peng10}
{Peng}, Y.-j., {Lilly}, S.~J., {Kova{\v{c}}}, K., {et~al.} 2010, \apj, 721, 193

\bibitem[{{P{\'e}roux} \& {Howk}(2020)}]{peroux20}
{P{\'e}roux}, C. \& {Howk}, J.~C. 2020, \araa, 58, 363

\bibitem[{{Postman} {et~al.}(2012){Postman}, {Coe}, {Ben{\'\i}tez}, {Bradley},
  {Broadhurst}, {Donahue}, {Ford}, {Graur}, {Graves}, {Jouvel}, {Koekemoer},
  {Lemze}, {Medezinski}, {Molino}, {Moustakas}, {Ogaz}, {Riess}, {Rodney},
  {Rosati}, {Umetsu}, {Zheng}, {Zitrin}, {Bartelmann}, {Bouwens}, {Czakon},
  {Golwala}, {Host}, {Infante}, {Jha}, {Jimenez-Teja}, {Kelson}, {Lahav},
  {Lazkoz}, {Maoz}, {McCully}, {Melchior}, {Meneghetti}, {Merten}, {Moustakas},
  {Nonino}, {Patel}, {Reg{\"o}s}, {Sayers}, {Seitz}, \& {Van der
  Wel}}]{postman12}
{Postman}, M., {Coe}, D., {Ben{\'\i}tez}, N., {et~al.} 2012, \apjs, 199, 25

\bibitem[{{R{\'e}my-Ruyer} {et~al.}(2014){R{\'e}my-Ruyer}, {Madden},
  {Galliano}, {Galametz}, {Takeuchi}, {Asano}, {Zhukovska}, {Lebouteiller},
  {Cormier}, {Jones}, {Bocchio}, {Baes}, {Bendo}, {Boquien}, {Boselli},
  {DeLooze}, {Doublier-Pritchard}, {Hughes}, {Karczewski}, \&
  {Spinoglio}}]{remy14}
{R{\'e}my-Ruyer}, A., {Madden}, S.~C., {Galliano}, F., {et~al.} 2014, \aap,
  563, A31

\bibitem[{{Riechers} {et~al.}(2019){Riechers}, {Pavesi}, {Sharon}, {Hodge},
  {Decarli}, {Walter}, {Carilli}, {Aravena}, {da Cunha}, {Daddi}, {Dickinson},
  {Smail}, {Capak}, {Ivison}, {Sargent}, {Scoville}, \& {Wagg}}]{riechers19}
{Riechers}, D.~A., {Pavesi}, R., {Sharon}, C.~E., {et~al.} 2019, \apj, 872, 7

\bibitem[{{Rodighiero} {et~al.}(2010){Rodighiero}, {Cimatti}, {Gruppioni},
  {Popesso}, {Andreani}, {Altieri}, {Aussel}, {Berta}, {Bongiovanni},
  {Brisbin}, {Cava}, {Cepa}, {Daddi}, {Dominguez-Sanchez}, {Elbaz}, {Fontana},
  {F{\"o}rster Schreiber}, {Franceschini}, {Genzel}, {Grazian}, {Lutz},
  {Magdis}, {Magliocchetti}, {Magnelli}, {Maiolino}, {Mancini}, {Nordon},
  {Perez Garcia}, {Poglitsch}, {Santini}, {Sanchez-Portal}, {Pozzi},
  {Riguccini}, {Saintonge}, {Shao}, {Sturm}, {Tacconi}, {Valtchanov},
  {Wetzstein}, \& {Wieprecht}}]{rodighiero10}
{Rodighiero}, G., {Cimatti}, A., {Gruppioni}, C., {et~al.} 2010, \aap, 518, L25

\bibitem[{{Sargent} {et~al.}(2014){Sargent}, {Daddi}, {B{\'e}thermin},
  {Aussel}, {Magdis}, {Hwang}, {Juneau}, {Elbaz}, \& {da Cunha}}]{sargent14}
{Sargent}, M.~T., {Daddi}, E., {B{\'e}thermin}, M., {et~al.} 2014, \apj, 793,
  19

\bibitem[{{Schouws} {et~al.}(2022{\natexlab{a}}){Schouws}, {Stefanon},
  {Bouwens}, \& al}]{schouws22b}
{Schouws}, S., {Stefanon}, M., {Bouwens}, R., \& al, e. 2022{\natexlab{a}},
  \apj, 999, 9

\bibitem[{{Schouws} {et~al.}(2022{\natexlab{b}}){Schouws}, {Stefanon},
  {Bouwens}, {Smit}, {Hodge}, {Labb{\'e}}, {Algera}, {Boogaard}, {Carniani},
  {Fudamoto}, {Holwerda}, {Illingworth}, {Maiolino}, {Maseda}, {Oesch}, \& {van
  der Werf}}]{schouws22a}
{Schouws}, S., {Stefanon}, M., {Bouwens}, R., {et~al.} 2022{\natexlab{b}},
  \apj, 928, 31

\bibitem[{{Schreiber} {et~al.}(2015){Schreiber}, {Pannella}, {Elbaz},
  {B{\'e}thermin}, {Inami}, {Dickinson}, {Magnelli}, {Wang}, {Aussel}, {Daddi},
  {Juneau}, {Shu}, {Sargent}, {Buat}, {Faber}, {Ferguson}, {Giavalisco},
  {Koekemoer}, {Magdis}, {Morrison}, {Papovich}, {Santini}, \&
  {Scott}}]{schreiber15}
{Schreiber}, C., {Pannella}, M., {Elbaz}, D., {et~al.} 2015, \aap, 575, A74

\bibitem[{{Scoville} {et~al.}(2007){Scoville}, {Abraham}, {Aussel}, {Barnes},
  {Benson}, {Blain}, {Calzetti}, {Comastri}, {Capak}, {Carilli}, {Carlstrom},
  {Carollo}, {Colbert}, {Daddi}, {Ellis}, {Elvis}, {Ewald}, {Fall},
  {Franceschini}, {Giavalisco}, {Green}, {Griffiths}, {Guzzo}, {Hasinger},
  {Impey}, {Kneib}, {Koda}, {Koekemoer}, {Lefevre}, {Lilly}, {Liu},
  {McCracken}, {Massey}, {Mellier}, {Miyazaki}, {Mobasher}, {Mould}, {Norman},
  {Refregier}, {Renzini}, {Rhodes}, {Rich}, {Sanders}, {Schiminovich},
  {Schinnerer}, {Scodeggio}, {Sheth}, {Shopbell}, {Taniguchi}, {Tyson}, {Urry},
  {Van Waerbeke}, {Vettolani}, {White}, \& {Yan}}]{scoville07}
{Scoville}, N., {Abraham}, R.~G., {Aussel}, H., {et~al.} 2007, \apjs, 172, 38

\bibitem[{{Scoville} {et~al.}(2014){Scoville}, {Aussel}, {Sheth}, {Scott},
  {Sanders}, {Ivison}, {Pope}, {Capak}, {Vanden Bout}, {Manohar}, {Kartaltepe},
  {Robertson}, \& {Lilly}}]{scoville14}
{Scoville}, N., {Aussel}, H., {Sheth}, K., {et~al.} 2014, \apj, 783, 84

\bibitem[{{Scoville} {et~al.}(2017){Scoville}, {Lee}, {Vanden Bout},
  {Diaz-Santos}, {Sanders}, {Darvish}, {Bongiorno}, {Casey}, {Murchikova},
  {Koda}, {Capak}, {Vlahakis}, {Ilbert}, {Sheth}, {Morokuma-Matsui}, {Ivison},
  {Aussel}, {Laigle}, {McCracken}, {Armus}, {Pope}, {Toft}, \&
  {Masters}}]{scoville17}
{Scoville}, N., {Lee}, N., {Vanden Bout}, P., {et~al.} 2017, \apj, 837, 150

\bibitem[{{Smit} {et~al.}(2018){Smit}, {Bouwens}, {Carniani}, {Oesch},
  {Labb{\'e}}, {Illingworth}, {van der Werf}, {Bradley}, {Gonzalez}, {Hodge},
  {Holwerda}, {Maiolino}, \& {Zheng}}]{smit18}
{Smit}, R., {Bouwens}, R.~J., {Carniani}, S., {et~al.} 2018, \nat, 553, 178

\bibitem[{{Sommovigo} {et~al.}(2021){Sommovigo}, {Ferrara}, {Carniani},
  {Zanella}, {Pallottini}, {Gallerani}, \& {Vallini}}]{sommovigo21}
{Sommovigo}, L., {Ferrara}, A., {Carniani}, S., {et~al.} 2021, \mnras, 503,
  4878

\bibitem[{{Sommovigo} {et~al.}(2022){Sommovigo}, {Ferrara}, {Pallottini},
  {Dayal}, {Bouwens}, {Smit}, {da Cunha}, {De Looze}, {Bowler}, {Hodge},
  {Inami}, {Oesch}, {Endsley}, {Gonzalez}, {Schouws}, {Stark}, {Stefanon},
  {Aravena}, {Graziani}, {Riechers}, {Schneider}, {van der Werf}, {Algera},
  {Barrufet}, {Fudamoto}, {Hygate}, {Labb{\'e}}, {Li}, {Nanayakkara}, \&
  {Topping}}]{sommovigo22}
{Sommovigo}, L., {Ferrara}, A., {Pallottini}, A., {et~al.} 2022, \mnras, 513,
  3122

\bibitem[{{Speagle} {et~al.}(2014){Speagle}, {Steinhardt}, {Capak}, \&
  {Silverman}}]{speagle14}
{Speagle}, J.~S., {Steinhardt}, C.~L., {Capak}, P.~L., \& {Silverman}, J.~D.
  2014, \apjs, 214, 15

\bibitem[{{Spilker} {et~al.}(2018){Spilker}, {Bezanson}, {Bari{\v{s}}i{\'c}},
  {Bell}, {Lagos}, {Maseda}, {Muzzin}, {Pacifici}, {Sobral}, {Straatman}, {van
  der Wel}, {van Dokkum}, {Weiner}, {Whitaker}, {Williams}, \&
  {Wu}}]{spilker18}
{Spilker}, J., {Bezanson}, R., {Bari{\v{s}}i{\'c}}, I., {et~al.} 2018, \apj,
  860, 103

\bibitem[{{Tacconi} {et~al.}(2018){Tacconi}, {Genzel}, {Saintonge}, {Combes},
  {Garc{\'\i}a-Burillo}, {Neri}, {Bolatto}, {Contini}, {F{\"o}rster Schreiber},
  {Lilly}, {Lutz}, {Wuyts}, {Accurso}, {Boissier}, {Boone}, {Bouch{\'e}},
  {Bournaud}, {Burkert}, {Carollo}, {Cooper}, {Cox}, {Feruglio}, {Freundlich},
  {Herrera-Camus}, {Juneau}, {Lippa}, {Naab}, {Renzini}, {Salome}, {Sternberg},
  {Tadaki}, {{\"U}bler}, {Walter}, {Weiner}, \& {Weiss}}]{tacconi18}
{Tacconi}, L.~J., {Genzel}, R., {Saintonge}, A., {et~al.} 2018, \apj, 853, 179

\bibitem[{{Tacconi} {et~al.}(2020){Tacconi}, {Genzel}, \&
  {Sternberg}}]{tacconi20}
{Tacconi}, L.~J., {Genzel}, R., \& {Sternberg}, A. 2020, \araa, 58, 157

\bibitem[{{Tacconi} {et~al.}(2013){Tacconi}, {Neri}, {Genzel}, {Combes},
  {Bolatto}, {Cooper}, {Wuyts}, {Bournaud}, {Burkert}, {Comerford}, {Cox},
  {Davis}, {F{\"o}rster Schreiber}, {Garc{\'\i}a-Burillo}, {Gracia-Carpio},
  {Lutz}, {Naab}, {Newman}, {Omont}, {Saintonge}, {Shapiro Griffin}, {Shapley},
  {Sternberg}, \& {Weiner}}]{tacconi13}
{Tacconi}, L.~J., {Neri}, R., {Genzel}, R., {et~al.} 2013, \apj, 768, 74

\bibitem[{{Topping} {et~al.}(2022){Topping}, {Stark}, {Endsley}, {Bouwens},
  {Schouws}, {Smit}, {Stefanon}, {Inami}, {Bowler}, {Oesch}, {Gonzalez},
  {Dayal}, {da Cunha}, {Algera}, {van der Werf}, {Pallottini}, {Barrufet},
  {Schneider}, {De Looze}, {Sommovigo}, {Whitler}, {Graziani}, {Fudamoto}, \&
  {Ferrara}}]{topping22}
{Topping}, M.~W., {Stark}, D.~P., {Endsley}, R., {et~al.} 2022, \mnras, 516,
  975

\bibitem[{{Trenti} {et~al.}(2011){Trenti}, {Bradley}, {Stiavelli}, {Oesch},
  {Treu}, {Bouwens}, {Shull}, {MacKenty}, {Carollo}, \&
  {Illingworth}}]{trenti11}
{Trenti}, M., {Bradley}, L.~D., {Stiavelli}, M., {et~al.} 2011, \apjl, 727, L39

\bibitem[{{Valentino} {et~al.}(2018){Valentino}, {Magdis}, {Daddi}, {Liu},
  {Aravena}, {Bournaud}, {Cibinel}, {Cormier}, {Dickinson}, {Gao}, {Jin},
  {Juneau}, {Kartaltepe}, {Lee}, {Madden}, {Puglisi}, {Sanders}, \&
  {Silverman}}]{valentino18}
{Valentino}, F., {Magdis}, G.~E., {Daddi}, E., {et~al.} 2018, \apj, 869, 27

\bibitem[{{Valentino} {et~al.}(2020){Valentino}, {Magdis}, {Daddi}, {Liu},
  {Aravena}, {Bournaud}, {Cortzen}, {Gao}, {Jin}, {Juneau}, {Kartaltepe},
  {Kokorev}, {Lee}, {Madden}, {Narayanan}, {Popping}, \&
  {Puglisi}}]{valentino20}
{Valentino}, F., {Magdis}, G.~E., {Daddi}, E., {et~al.} 2020, \apj, 890, 24

\bibitem[{{Vallini} {et~al.}(2015){Vallini}, {Gallerani}, {Ferrara},
  {Pallottini}, \& {Yue}}]{vallini15}
{Vallini}, L., {Gallerani}, S., {Ferrara}, A., {Pallottini}, A., \& {Yue}, B.
  2015, \apj, 813, 36

\bibitem[{{Vizgan} {et~al.}(2022){Vizgan}, {Greve}, {Olsen}, {Zanella},
  {Narayanan}, {Dav{\`e}}, {Magdis}, {Popping}, {Valentino}, \&
  {Heintz}}]{vizgan22}
{Vizgan}, D., {Greve}, T.~R., {Olsen}, K.~P., {et~al.} 2022, \apj, 929, 92

\bibitem[{{Walter} {et~al.}(2020){Walter}, {Carilli}, {Neeleman}, {Decarli},
  {Popping}, {Somerville}, {Aravena}, {Bertoldi}, {Boogaard}, {Cox}, {da
  Cunha}, {Magnelli}, {Obreschkow}, {Riechers}, {Rix}, {Smail}, {Weiss},
  {Assef}, {Bauer}, {Bouwens}, {Contini}, {Cortes}, {Daddi}, {Diaz-Santos},
  {Gonz{\'a}lez-L{\'o}pez}, {Hennawi}, {Hodge}, {Inami}, {Ivison}, {Oesch},
  {Sargent}, {van der Werf}, {Wagg}, \& {Yung}}]{walter20}
{Walter}, F., {Carilli}, C., {Neeleman}, M., {et~al.} 2020, \apj, 902, 111

\bibitem[{{Walter} {et~al.}(2016){Walter}, {Decarli}, {Aravena}, {Carilli},
  {Bouwens}, {da Cunha}, {Daddi}, {Ivison}, {Riechers}, {Smail}, {Swinbank},
  {Weiss}, {Anguita}, {Assef}, {Bacon}, {Bauer}, {Bell}, {Bertoldi}, {Chapman},
  {Colina}, {Cortes}, {Cox}, {Dickinson}, {Elbaz}, {G{\'o}nzalez-L{\'o}pez},
  {Ibar}, {Inami}, {Infante}, {Hodge}, {Karim}, {Le Fevre}, {Magnelli}, {Neri},
  {Oesch}, {Ota}, {Popping}, {Rix}, {Sargent}, {Sheth}, {van der Wel}, {van der
  Werf}, \& {Wagg}}]{walter16}
{Walter}, F., {Decarli}, R., {Aravena}, M., {et~al.} 2016, \apj, 833, 67

\bibitem[{{Walter} {et~al.}(2014){Walter}, {Decarli}, {Sargent}, {Carilli},
  {Dickinson}, {Riechers}, {Ellis}, {Stark}, {Weiner}, {Aravena}, {Bell},
  {Bertoldi}, {Cox}, {Da Cunha}, {Daddi}, {Downes}, {Lentati}, {Maiolino},
  {Menten}, {Neri}, {Rix}, \& {Weiss}}]{walter14}
{Walter}, F., {Decarli}, R., {Sargent}, M., {et~al.} 2014, \apj, 782, 79

\bibitem[{{Whitaker} {et~al.}(2014){Whitaker}, {Franx}, {Leja}, {van Dokkum},
  {Henry}, {Skelton}, {Fumagalli}, {Momcheva}, {Brammer}, {Labb{\'e}},
  {Nelson}, \& {Rigby}}]{whitaker14}
{Whitaker}, K.~E., {Franx}, M., {Leja}, J., {et~al.} 2014, \apj, 795, 104

\bibitem[{{Whitaker} {et~al.}(2010){Whitaker}, {van Dokkum}, {Brammer},
  {Kriek}, {Franx}, {Labb{\'e}}, {Marchesini}, {Quadri}, {Bezanson},
  {Illingworth}, {Lee}, {Muzzin}, {Rudnick}, \& {Wake}}]{whitaker10}
{Whitaker}, K.~E., {van Dokkum}, P.~G., {Brammer}, G., {et~al.} 2010, \apj,
  719, 1715

\bibitem[{{Williams} {et~al.}(2021){Williams}, {Spilker}, {Whitaker},
  {Dav{\'e}}, {Woodrum}, {Brammer}, {Bezanson}, {Narayanan}, \&
  {Weiner}}]{williams21}
{Williams}, C.~C., {Spilker}, J.~S., {Whitaker}, K.~E., {et~al.} 2021, \apj,
  908, 54

\bibitem[{{Yan} {et~al.}(2011){Yan}, {Yan}, {Zamojski}, {Windhorst},
  {McCarthy}, {Fan}, {R{\"o}ttgering}, {Koekemoer}, {Robertson}, {Dav{\'e}}, \&
  {Cai}}]{yan11}
{Yan}, H., {Yan}, L., {Zamojski}, M.~A., {et~al.} 2011, \apjl, 728, L22

\bibitem[{{Yan} {et~al.}(2020){Yan}, {Sajina}, {Loiacono}, {Lagache},
  {B{\'e}thermin}, {Faisst}, {Ginolfi}, {F{\`e}vre}, {Gruppioni}, {Capak},
  {Cassata}, {Schaerer}, {Silverman}, {Bardelli}, {Dessauges-Zavadsky},
  {Cimatti}, {Hathi}, {Lemaux}, {Ibar}, {Jones}, {Koekemoer}, {Oesch}, {Talia},
  {Pozzi}, {Riechers}, {Tasca}, {Toft}, {Vallini}, {Vergani}, {Zamorani}, \&
  {Zucca}}]{yan20}
{Yan}, L., {Sajina}, A., {Loiacono}, F., {et~al.} 2020, \apj, 905, 147

\bibitem[{{Zanella} {et~al.}(2018){Zanella}, {Daddi}, {Magdis}, {Diaz Santos},
  {Cormier}, {Liu}, {Cibinel}, {Gobat}, {Dickinson}, {Sargent}, {Popping},
  {Madden}, {Bethermin}, {Hughes}, {Valentino}, {Rujopakarn}, {Pannella},
  {Bournaud}, {Walter}, {Wang}, {Elbaz}, \& {Coogan}}]{zanella18}
{Zanella}, A., {Daddi}, E., {Magdis}, G., {et~al.} 2018, \mnras, 481, 1976

\end{thebibliography}
%

\end{document}